\documentclass[a4paper,twocolumn,11pt]{iopart}
\makeatletter
\renewcommand\@appendixstar{\@@par
 \ifnumbysec 
 \@addtoreset{table}{section}
 \@addtoreset{figure}{section}\fi
 \setcounter{section}{0}
 \setcounter{subsection}{0}
 \setcounter{subsubsection}{0}
 \setcounter{equation}{0}
 \setcounter{figure}{0}
 \setcounter{table}{0}
 \def\thesection{\Alph{section}} % this line has been \def\thesection{Appendix \Alph{section}} before
 \def\theequation{\ifnumbysec
      \Alph{section}.\arabic{equation}\else
      \Alph{section}\arabic{equation}\fi}
 \def\thetable{\ifnumbysec
      \Alph{section}\arabic{table}\else
      A\arabic{table}\fi}
 \def\thefigure{\ifnumbysec
      \Alph{section}\arabic{figure}\else
      A\arabic{figure}\fi}}
\makeatother

\usepackage{iopams} 
\expandafter\let\csname equation*\endcsname\relax

\expandafter\let\csname endequation*\endcsname\relax
\pdfoutput=1
\usepackage[utf8]{inputenc}
\usepackage[english]{babel}
\usepackage[T1]{fontenc}
\usepackage{amsmath,amsfonts,enumerate,amsthm,amssymb,mathtools}
\usepackage{hyperref}

\usepackage{bm}
\usepackage{braket}
\usepackage{amsmath}
\usepackage{comment}

\usepackage{xcolor}

\usepackage{caption}
\usepackage{subcaption}

\newtheorem{thm}{Theorem}
\newtheorem{lem}[thm]{Lemma}
\newtheorem{defn}{Definition}

\newcommand{\kket}[1]{|#1 \rangle \rangle}
\newcommand{\bbra}[1]{ \langle \langle #1 |}
\newcommand{\bbrakket}[2]{\langle \langle #1 | #2 \rangle \rangle}
\newcommand{\pw}[1]{P({#1})}
\newcommand{\pwn}[1]{\overline{P}({#1})}

\begin{document}

\title{Emergence of noise-induced barren plateaus in arbitrary layered noise models}
\author{M. Schumann$^{1, 2}$\footnote{ma.schumann@fz-juelich.de}, F. K. Wilhelm$^{1,2}$, A. Ciani$^1$}
\address{$^1$ Institute for Quantum Computing Analytics (PGI-12), Forschungszentrum J\"ulich, 52425 J\"ulich, Germany}
\address{$^2$ Theoretical Physics, Saarland University, 66123 Saarbr\"ucken, Germany}

\maketitle

\begin{abstract}
In variational quantum algorithms the parameters of a parameterized quantum circuit are optimized in order to minimize a cost function that encodes the solution of the problem. The barren plateau phenomenon manifests as an exponentially vanishing dependence of the cost function with respect to the variational parameters, and thus hampers the optimization process. We discuss how, and in which sense, the phenomenon of noise-induced barren plateaus emerges in parameterized quantum circuits with a layered noise model. Previous results have shown the existence of noise-induced barren plateaus in the presence of local Pauli noise \cite{Wang}. We extend these results analytically to arbitrary completely-positive trace preserving maps  in two cases: 1) when a parameter-shift rule holds, 2) when the parameterized quantum circuit at each layer forms a unitary $2$-design. The second example shows how highly expressive unitaries give rise not only to standard barren plateaus \cite{McClean2018}, but also to noise-induced ones. In the second part of the paper, we study numerically the emergence of noise-induced barren plateaus in QAOA circuits focusing on the case of MaxCut problems on $d$-regular graphs and amplitude damping noise. 
\end{abstract}

\section{Introduction}

The last decade has witnessed a significant development of quantum computing hardware in several experimental platforms, such as those based on superconducting and trapped ions qubits. Despite the increase in the number of simultaneously controllable qubits on a chip, which has now passed the $100$-qubit barrier, current error rates are still significant. As a consequence, current devices do not allow to fully exploit the power of quantum computing needed to run algorithms with an exponential speed-up over the best known classical methods, such as the celebrated Shor's algorithm \cite{shor1997, Gidney2021}. Quantum error correction \cite{terhal2015} holds the promise that, as long as the error rates are below a certain threshold value, the logical errors can be brought to an arbitrarily low level. However, current experimental demonstrations of quantum error correction based on the surface code \cite{fowler2012, dennis2002} have not yet convincingly demonstrated the power of error correction, although the results are promising \cite{googleqec, Krinner2022, Marques2022}. 

The current quantum computers where the number of qubits is large, but errors are still dominant, have been nicknamed Noisy-Intermediate Scale Quantum (NISQ) devices \cite{preskill2018}. Variational quantum algorithms (VQAs) \cite{Cerezo2021} have been the main focus in the current NISQ era. In VQAs, a parameterized quantum circuit (PQC) is used to evaluate a cost function that encodes the solution of a given problem. The evaluation of the cost function is supposed to be the \emph{hard} part of the algorithm and thus delegated to the quantum computer. The parameters in the quantum circuit are optimized classically via standard methods in order to minimize the cost function. One says that the parameters are \emph{trained} in analogy with neural networks in classical machine learning. This basic hybrid quantum-classical routine is the basis for VQAs such as the variational quantum eigensolvers \cite{Peruzzo2014, TILLY20221} and the quantum approximate optimization algorithm (QAOA) \cite{farhi2014, zhou2020, hadfield2019}, as well as variational algorithms for linear algebra \cite{xu2021, bravoprieto2019} and differential equations \cite{kyriienko2021, Leong2022, kubo2021}. Additionally, PQCs are often considered as a the quantum analog of classical neural networks \cite{mitarai2018, Abbas2021}. 

In order for a VQA to be successful at least two conditions must be met. First, there must exist parameters that give a solution that is sufficiently close to the ideal one. We remark that it is usually hard to obtain performance guarantees in VQA, and as such they are considered heuristic methods. Second, we must be able to find such parameters. In this case, one says that the system has to be trainable. After the first proposals for VQAs, it was soon realized that a high ansatz expressivity, while desirable to guarantee the existence of a good solution, is at odds with the trainability of PQCs. In particular, Ref.~\cite{McClean2018} showed that when the PQC forms a unitary $2$-design \cite{emerson2005, dankert2009}, the variance of the gradient of a cost function evaluated with the PQC vanishes exponentially with the number of qubits in the system. We refer to this phenomenon as the standard barren plateau (SBP) phenomenon. A number of follow-up works have further investigated the connection between expressivity and trainability \cite{holmes2022, Friedrich2023, Larocca2022, ortiz2021, Zhao2021analyzingbarren, Lubasch_2023}. The barren plateau phenomenon clearly affects training via gradient-based methods, but they also affect gradient-free optimization \cite{arrasmith2021, Arrasmith_2022}.  

Recently, Ref.~\cite{Wang} has pointed out that, in the presence of noise, barren plateaus can manifest as an exponential decay of the gradient of the cost function with the number of layers in the quantum circuit. This phenomenon is known as noise-induced barren plateau (NIBP). In particular, Ref.~\cite{Wang} showed the results for a layered, single-qubit Pauli noise model. We remark that a different line of research has focused on the limitations of VQAs algorithm in the presence of noise, independently of the barren plateau problem \cite{StilckFrança2021, depalma2023, skolik2022, Xue_2021, Marshall_2020, cirac2022}.

In this work, we study the NIBP phenomenon for more general noise maps, extending the results of Ref.~\cite{Wang}. In particular, we consider a Markovian, layered noise model shown in Fig.~\ref{fig:toy_model}, where parameterized unitaries are alternated with a noise map $\mathcal{N}$, that is a completely positive, trace-preserving (CPTP) map. We provide analytical derivations, as well as numerical results for small instances of QAOA circuits for MaxCut problems on $d$-regular graphs in the presence of single-qubit amplitude damping noise.  

The paper is organized as follows. Sec.~\ref{sec:toy_model} introduces the error model and sets up the notation. Further preliminary concepts are provided in Appendix~\ref{app:preliminary}. In Sec.~\ref{sec:analytical} we summarize our analytical results. Details of the analytical derivations are gathered in Appendix~\ref{app:var_grad_decay} and \ref{sec:app_c}. In Sec.~\ref{sec:numerical} we discuss our numerical results for QAOA circuits with amplitude damping noise. Additional numerical results can be found in Appendix~\ref{sec:additional_qaoa}. We draw the conclusions and provide an outlook in Sec.~\ref{sec:conclusion}.

\section{Description of the error model}
\label{sec:toy_model}

\begin{figure}
    \centering
    \includegraphics[width=1\textwidth]{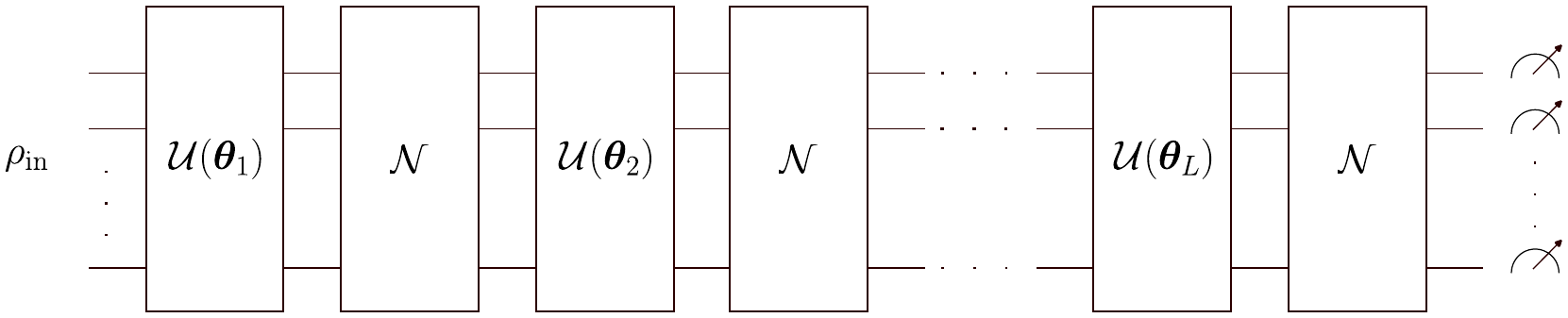}
    \caption{The Markovian, layered error model we consider in our analysis. The model consists of $L$ alternating parameterized channels $\mathcal{U}(\bm{\theta}_{\ell})$  that correspond to the parameterized unitaries $U(\bm{\theta}_{\ell})$. They are followed by a CPTP noise map $\mathcal{N}$. }
    \label{fig:toy_model}
\end{figure}

In what follows, we consider a quantum mechanical system composed of $n$ qubits. We denote the quantum operation associated with a unitary $U$ as $\mathcal{U}$ that acts on a density matrix $\rho$ as $\mathcal{U}(\rho) = U \rho U^{\dagger}$. The action of a CPTP map $\mathcal{N}$ on a density matrix $\rho$ can always be written as $\mathcal{N}(\rho) = \sum_j A_j \rho A_j^{\dagger}$, with $A_j$ the Kraus operators satisfying $\sum_j A_j^{\dagger} A_j = I$  and $I$ the identity on $n$ qubits. We will denote by $\hat{\mathcal{N}}$ the Liouville superoperators associated with the noise map $\mathcal{N}$. The representation of $\hat{\mathcal{N}}$ in the $n$-qubit Pauli basis is known as the Pauli Transfer Matrix (PTM) (see Appendix~\ref{app:preliminary} for more details). Note that $\hat{\mathcal{N}}$ is represented as a $4^n \times 4^n$ matrix.

Similar to previous literature \cite{ Wang,fontana2021, wang2021, StilckFrança2021}, we consider a layered noise model that consists of alternating applications of parameterized unitaries and a CPTP map $\mathcal{N}$, as shown in Fig.~\ref{fig:toy_model}. This is a Markovian noise model, since the state of the system after the application of the noise map $\mathcal{N}$ depends only on the state before. However, the noise map itself can be a generic CPTP map that admits a Kraus decomposition, and does not need to form necessarily a semigroup of the form $\mathcal{N}(t)= e^{t \mathcal{L}}$ with $t>0$ and $\mathcal{L}$ a generator in Lindblad form \cite{petruccione, manzano2020}. We refer to these kinds of Markovian maps as Lindbladian noise maps, which prominently include the case of single-qubit amplitude damping and dephasing noise. 

We assume that the unitary at layer $\ell$ can be written as $U(\bm{\theta}_{\ell})$, i.e., it is parameterized by a set of parameters that we gather in a vector $\bm{\theta}_{\ell} = \begin{pmatrix} \theta_{\ell 1} & \dots & \theta_{\ell K} \end{pmatrix}$, where $\ell = 1, \dots, L$ with $L$ the total number of layers and $K$ the number of parameters per layer. For simplicity, we assume that each layer has the same number of parameters and that the unitaries have the same structure at each layer\footnote{One can also consider the scenario where each layers has a different number of parameters and unitaries with different structure and obtain analogous results, at the price of a more tedious derivation.}. Thus, we can always write the parameterized unitary as

\begin{equation}
    U(\bm{\theta}_{\ell}) = \prod_{k=1}^K U_{k}(\theta_{\ell k}).
\end{equation}

 Circuits of this form are said to have a periodic structure \cite{Larocca2022}. We denote by $\mathcal{E}(\bm{\theta})$ the quantum operation associated with the circuit in Fig.~\ref{fig:toy_model}, that is
 
\begin{equation}
    \mathcal{E} (\bm{\theta}) = \mathcal{N} \circ \mathcal{U}(\bm{\theta}_L) \circ \dots \mathcal{N} \circ \mathcal{U}(\bm{\theta}_1),
\end{equation}

\noindent where $\bm{\theta} = \bigoplus_{\ell=1}^L \bm{\theta}_{\ell}$ is the vector of all parameters. Assuming an initial state $\rho_{\mathrm{in}}$, the parameterized state at the output of the quantum circuit is  

\begin{equation}
\rho(\bm{\theta}) = \mathcal{E}(\bm{\theta})(\rho_{\mathrm{in}}). 
\end{equation}

In VQAs we are interested in minimizing (or maximizing) a cost function that is the average of an observable $O$ on the output state of a PQC. Without loss of generality, we can take $O$ to be Hermitian and traceless. The noisy cost function is given by

\begin{equation}
\label{eq:costf}
    C(\bm{\theta}) = \mathrm{Tr} [ O \rho(\bm{\theta}) ].
\end{equation}

\section{Analytical results}
\label{sec:analytical}

The barren plateau phenomenon in VQAs manifests itself as a flat cost function landscape. This phenomenon can be exponential in the number of qubits, as in the standard formulation of barren plateaus \cite{McClean2018}, or in the number of layers, as it happens in the case of NIBPs \cite{Wang}.  

In the literature, SBPs have been mostly characterized via the decay of the variance of the partial derivatives when the parameters are sampled according to a certain probability distribution \cite{McClean2018, Larocca2022}. In fact, if the variance is small, and the average is zero, we would sample with high probability initial parameters that give very small gradients, and thus render the training of the parameters unfeasible. On the other hand, NIBPs have been studied for noise models, such as single-qubit Pauli noise \cite{Wang}, that give a somewhat stronger notion of barren plateaus, where the cost function and the partial derivatives vanish exponentially with the number of layers for $\emph{any}$ parameter. 

In what follows, we will extend the types of noise models that give rise to NIBP. In particular, we will study in which sense NIBP emerge in two special cases:

\begin{enumerate}
    \item The partial derivative can be evaluated via a parameter-shift rule;
\item The parameterized unitaries at each layer form a unitary $2$-design.
\end{enumerate}

\subsection{NIBP with parameter-shift rule}
\label{subsec:nibpshift}

Let us suppose that a parameter-shift rule holds for the parameter $\theta_{\ell k}$ \cite{mitarai2018, schuld2018, crooks2019}. This means that the partial derivative with respect to $\theta_{\ell k}$ can be evaluated as the difference of the cost function at two different points in parameter space:

\begin{equation}
\label{eq:psrule}
    \frac{\partial C}{\partial \theta_{\ell k}} = C(\bm{\theta} + \alpha \bm{e}_{\ell k}) - C(\bm{\theta} + \beta \bm{e}_{\ell k}),
\end{equation}

\noindent where $\alpha, \beta \in \mathbb{R}$ and we denoted by $\bm{e}_{\ell k}$ the unit vector that has components $(\bm{e}_{\ell k})_{\ell' k'}= \delta_{\ell \ell'} \delta_{k k'}$. A prominent example where the parameter-shift rule holds is in the case of parameterized unitaries of the form $e^{-i \theta_{\ell k} P}$ with $P$ a Pauli operator, for which $\alpha = -\beta = \pi/4$ \cite{crooks2019}. Notice that in many proposals for VQAs, for instance those based on a hardware efficient ansatz \cite{Kandala2017} or in multi-angle QAOA \cite{Herrman2022}, there are parameterized unitaries of this form. 

Looking at Eq.~\eqref{eq:psrule}, we see that when a parameter-shift rule holds, the concentration of the cost function implies that the partial derivative vanishes. Let us now argue that a large class of CPTP maps $\mathcal{N}$ give rise to NIBP in this setting. Let $\rho_{\ell k}^{(\alpha)}$ and $\rho_{\ell k}^{(\beta)}$ be the density matrices obtained after the application of the quantum circuit up to the channels $\mathcal{U}_{ k}(\theta_{\ell k} + \alpha)$, $\mathcal{U}_{ k}(\theta_{\ell k} + \beta)$, respectively:

\begin{subequations}
\begin{equation}
    \rho_{\ell k}^{(\alpha)} = \mathcal{U}_{k}(\theta_{\ell k} + \alpha) \circ \dots \circ \mathcal{N} \circ \mathcal{U}(\bm{\theta}_{\ell -1}) \circ \dots   \circ \mathcal{N} \circ \mathcal{U}(\bm{\theta}_1) (\rho_{\mathrm{in}}),
\end{equation}
\begin{equation}
    \rho_{\ell k}^{(\beta)} = \mathcal{U}_{k}(\theta_{\ell k} + \beta) \circ \dots \circ \mathcal{N} \circ \mathcal{U}(\bm{\theta}_{\ell -1}) \circ \dots \circ \mathcal{N} \circ \mathcal{U}(\bm{\theta}_1) (\rho_{\mathrm{in}}).
\end{equation}
\end{subequations}

In order to compute the partial derivative, we need to evaluate the average of the observable $O$ on the two final states

\begin{subequations}
\begin{equation}
    \rho_{f}^{(\alpha)} = \mathcal{N} \circ \mathcal{U}(\bm{\theta}_L) \circ \dots \circ \mathcal{N} \circ \mathcal{U}_K(\theta_{\ell K}) \circ  \dots \circ \mathcal{U}_{k+1}(\theta_{\ell (k +1)})  \bigl(\rho_{\ell k}^{(\alpha)} \bigr), 
    \end{equation}
    \begin{equation}
    \rho_{f}^{(\beta)} = \mathcal{N} \circ \mathcal{U}(\bm{\theta}_L) \circ \dots \circ \mathcal{N} \circ \mathcal{U}_K(\theta_{\ell K}) \circ  \dots \circ \mathcal{U}_{k+1}(\theta_{\ell (k +1)})  \bigl(\rho_{\ell k}^{(\beta)} \bigr).
  \end{equation}
\end{subequations}

The density matrices $\rho_{\ell k}^{\alpha}$, $\rho_{\ell k}^{\beta}$ are evolved via the same quantum circuit of alternating unitaries and noise maps. A property of CPTP maps is that they make quantum states more indistinguishable. For instance, let us consider the trace distance between two quantum states $\rho_1$ and $\rho_2$ defined as

\begin{equation}
    T(\rho_1, \rho_2) = \frac{\lVert \rho_1 - \rho_2 \rVert_1}{2},
\end{equation}

\noindent with $\lVert \cdot \rVert_1$ the Schatten-$1$ norm (see Eq.~\eqref{eq:schatten} in the Appendix). It is known that for any CPTP map $\mathcal{N}$ \cite{NielsenChuang} 

\begin{equation}
\label{eq:ineqtd}
    T(\mathcal{N}(\rho_1), \mathcal{N}(\rho_2)) \le T(\rho_1, \rho_2).
\end{equation}

Additionally, equality holds for any $\rho_1, \rho_2$ if the map is unitary. Thus, for noise maps $\mathcal{N}$ for which the inequality in Eq.~\eqref{eq:ineqtd} is strictly decreasing, we expect that the repeated application of the map will make the states more and more indistinguishable. Let us define the contraction coefficient \cite{fumioRuskai2015, raginsky2002} or Lipschitz constant \cite{granasBook} of the CPTP map $\mathcal{N}$ as

\begin{equation}
    q_{\mathcal{N}} = \underbrace{\mathrm{sup}}_{\rho_1, \rho_2 \in \mathrm{D}_n, \rho_1 \neq \rho_2} \frac{ T(\mathcal{N}(\rho_1), \mathcal{N}(\rho_2))}{T(\rho_1, \rho_2)}.
\end{equation}

\noindent Clearly, $0 \le q_{\mathcal{N}} \le 1$ and thus
\begin{equation}
\label{eq:ineqtd2}
    T(\mathcal{N}(\rho_1), \mathcal{N}(\rho_2)) \le q_{\mathcal{N}} T(\rho_1, \rho_2).
\end{equation}

\noindent Using the matrix H\"{o}lder inequality Eq.~\eqref{eq:holders} in Appendix~\ref{app:preliminary} and iteratively using Eq.~\eqref{eq:ineqtd}, we get

\begin{equation}
\label{eq:nibpparshift}
    \biggl \lvert \frac{\partial C}{\partial \theta_{\ell k}} \biggr \rvert   = \Bigl \lvert \mathrm{Tr}\Bigl[O \Bigl(\rho_{f}^{(\alpha)} - \rho_{f}^{(\beta)} \Bigr) \Bigr]\Bigr \rvert \le
    2\lVert O \rVert_{\infty} T\Bigl(\rho_{f}^{(\alpha)}, \rho_{f}^{(\beta)} \Bigr) \le 
    2\lVert O \rVert_{\infty} q_{\mathcal{N}}^{L - \ell+1} T \Bigl( \rho_{\ell k}^{(\alpha)}, \rho_{\ell k}^{(\beta)} \Bigr). 
\end{equation}

We conclude that for strictly contractive noise maps $\mathcal{N}$ such that $q_{\mathcal{N}} <1$, if we keep the layer $\ell$ fixed and we increase the total number of layers $L$, we have an exponential decay of the partial derivative at \emph{any} point in parameter space. Qualitatively, this means that, if a parameter-shift rule holds for a certain parameter, it is not worth it to increase the total number of layers, since the system will exponentially forget its dependency on the parameter. Notice that this property does not rely on the channel being unital.

Strictly contractive noise maps are those that admit a unique fixed point and such that their repeated application leads arbitrarily close to the fixed point by virtue of Banach's contraction principle \cite{granasBook}. Prominent examples of strictly contractive noise channels are the single-qubit amplitude damping channel and the single-qubit Pauli channel with nonzero probability of having $X$, $Y$ and $Z$ errors. 

\subsection{NIBP with unitary $2$-designs}
\label{subsec:nibp2design}

We now move to analyze the case in which a parameter-shift rule does not necessarily hold, but similar to the original analysis of standard barren plateaus \cite{McClean2018}  the parameterized unitaries are \emph{sufficiently expressive}. In particular, we assume that when we randomly sample the parameters $\bm{\theta}_{\ell}$ in each layer, the unitaries form a unitary $2$-design \cite{emerson2005, dankert2009}. Mathematically, this means that for every noise map $\mathcal{N}$ it holds that

\begin{multline}
\label{eq:twirling_eq}
  \mathcal{N}_{\mathrm{Haar}}(\rho_{\mathrm{in}})  \coloneq
    \int d \mu(U) \mathcal{U}^{\dagger} \circ \mathcal{N} \circ \mathcal{U} (\rho_{\mathrm{in}})    
    =  \int_{\mathcal{D}} d \bm{\theta}' p(\bm{\theta}') \mathcal{U}^{\dagger}(\bm{\theta}') \circ \mathcal{N} \circ \mathcal{U}(\bm{\theta}') (\rho_{\mathrm{in}})   \\ 
    = (1-p_{\mathrm{eff}})\rho_{\mathrm{in}} +  p_{\mathrm{eff}} \frac{I}{2^n},
\end{multline}

\noindent where $d \mu (U)$ denotes the Haar random measure over the group $\mathrm{U}(2^n)$ of $2^n \times 2^n$ unitary matrices, $I$ the identity on $n$ qubits, $\mathcal{D}$ the domain of the parameters and $p(\bm{\theta}')$ the probability distribution with which we sample the parameters. The parameter $p_{\mathrm{eff}}$ of the global depolarizing channel appearing on the right-hand side of Eq.~\eqref{eq:twirling_eq} is given by \cite{emerson2005}

\begin{equation}
\label{eq:peffmaintext}
    p_{\mathrm{eff}} 
    = 1 - \frac{\sum_j | \mathrm{Tr}(A_{j}) |^2 -1}{2^{2n} -1}  
    = 1 - \frac{\mathrm{Tr}(\hat{\mathcal{N}}) - 1}{2^{2 n} - 1},
\end{equation}

\noindent with $A_{j}$ the Kraus operators of the noise map $\mathcal{N}$. We refer the reader to Appendix~\ref{subapp:basicgroup} for the basic representation theory needed to derive Eqs.~\eqref{eq:twirling_eq} and \eqref{eq:peffmaintext},  and to Ref.~\cite{mele2023} for a comprehensive discussion on unitary designs.

The toy model we are considering, without variational parameters, has been already used in Ref.\:\cite{eisert} to study the limitations of error mitigation protocols, and also in Ref.~\cite{bouland2022} to study random quantum circuits. Besides, as previously mentioned, Ref.~\cite{McClean2018} where SBPs in quantum circuits were first studied, also assumed quantum circuits with unitary $2$-design properties. Thus, our model falls into the same category as these contributions and allows to capture the limitations of error mitigation techniques, SBPs and NIBPs in a similar way. For these features to emerge it is sufficient, although not necessary, that the variational circuit is \emph{expressive enough} of the whole unitary group \cite{holmes2022}, which is a notion conveyed by the unitary $2$-design property. While the toy model is highly idealized, in Sec.~\ref{sec:numerical} we show via numerical simulations that it still allows to capture the behavior of QAOA circuits with local noise models, such as amplitude damping noise. 

In Appendix~\ref{app:var_grad_decay}, we obtain that for any noise map the variance of the derivative of any cost function in the toy model can be evaluated exactly and is given by 
\begin{equation}
\label{eq:var_decay}
 \mathrm{Var}\biggl( \frac{\partial C}{\partial \theta_{\ell k}} \biggr) = G_{\ell k} \frac{r_{\mathcal{N}}^{L - \ell}}{4^n-1} \mathrm{Tr}(\mathcal{N}^{\dagger}(O)^2),
 \end{equation}
 
\noindent where the coefficient $G_{\ell k}$ is given in Eq.~\eqref{eq:gkl} and can be evaluated iteratively, while the noise coefficient $r_{\mathcal{N}}$ in terms of Liouville superoperators reads
 
 \begin{equation}
     \label{eq:rn_maintext}
     r_{\mathcal{N}} = \frac{\mathrm{Tr}[ \hat{\mathcal{N}}^{\dagger} \hat{\mathcal{N}} (\hat{\mathcal{I}} - \kket{\pwn{0}}\bbra{\pwn{0}})]}{4^n-1}, 
 \end{equation}
 
\noindent with $\hat{\mathcal{I}}$ the $4^n \times 4^n$ identity matrix and $\kket{\pwn{0}}$ the vectorization of $I/\sqrt{2^{n}}$ (see Appendix~\ref{app:preliminary} for a summary of preliminary concepts). We remark that it was proven in Ref.~\cite{eisert} (see Proposition $2$ in Supplementary Material VII) that $0    <r_{\mathcal{N}} \le 1$ \footnote{Ref.~\cite{eisert} expressed the parameters in terms of the Choi state, while we define them in Liouville representation.}.  
 
 Let us comment on the interpretation of Eq.~\eqref{eq:var_decay}. We see that as long as $G_{\ell k} \mathrm{Tr}(\hat{\mathcal{N}}^{\dagger}(O)^2)$ does not increase as $\overline{q}^n$ with $\overline{q} \ge 4$ the system will show SBPs, i.e., an exponential decay of the variance with the number of qubits $n$, for all $\ell=1, \dots, L$, $k=1, \dots, K$. Regarding NIBPs, similar to the case of the parameter-shift rule, we see that if we fix the layer $\ell$ and increase the total number of layers $L$, we have an exponential decay with the total number of layers $L$. This happens for any noise map as long as $r_{\mathcal{N}}$ is strictly less than $1$. Thus, Eq.~\eqref{eq:var_decay} shows that high ansatz expressivity is not only expected to give rise to SBPs, but also to NIBPs in the previous sense, independently of the details of the noise map. However, let us consider the case in which we want to compute the derivative with respect to parameters in some of the last layers. In this case, if no further assumption on the noise is placed, we cannot claim that we have an exponential decay with the number of layers $L$. This can be understood with a simple example. Let us consider strong amplitude damping noise that will have the tendency to bring any state close to $\ket{0}^{\otimes n}$. We can assume that after a certain amount of layers, the system would end being approximately $\ket{0}^{\otimes n}$ if the noise is sufficiently strong. At this point, we will start again to apply random unitaries followed by noise starting from an approximately pure state, and so we expect that if we do not add too many additional layers we would preserve the dependency of the output on these parameters. 
 
 Eq.~\eqref{eq:var_decay} is valid for any noise map $\mathcal{N}$. We now analyze what happens in the case in which the map is unital, and when the map is is a Lindbladian noise map with weak noise strength. 
 A unital map, by definition, preserves the identity, i.e., $\mathcal{N}(I)=I$ and it is known that these maps are purity decreasing \cite{lidar2006}. In this case, Eq.~\eqref{eq:var_decay} becomes 

 \begin{equation}
     \label{eq:var_decay_unital} 
 \mathrm{Var}\biggl( \frac{\partial C}{\partial \theta_{\ell k}} \biggr) = \langle \nu_{\ell k}^{(-)} \rangle \frac{\mathrm{Tr}(\rho_{\mathrm{in}}^2) - 1/2^n}{(4^n -1)^2} 
 r_{\mathcal{N}}^{L - 2}  \mathrm{Tr}(\mathcal{N}^{\dagger}(O)^2).
 \end{equation}

\noindent This shows that for unital noise maps, NIBPs are expected for \emph{any} parameter differently than the general case we discussed before. 

 As we discuss in Appendix~\ref{subapp:markovian}, a purely dissipative Lindbladian noise map can always be written as the exponential of a Lindblad dissipator with jump operators $L_k$ \cite{petruccione}. In this case, and keeping only first order terms in the noise strength, the coefficient $r_{\mathcal{N}}$ in Eq.~\eqref{eq:rn_maintext} can be approximated as (see Appendix~\ref{app:var_grad_decay}) 

  \begin{equation}
 \label{eq:rnmarkovian}
     r_{\mathcal{N}} \approx \frac{4^n[1 -\frac{1}{2^{n-1}} \sum_{k} \mathrm{Tr}(L_k L_k^{\dagger})] -1}{4^n-1} 
     \overset{n \gg 1}{\approx} 1 -\frac{1}{2^{n-1}} \sum_{k} \mathrm{Tr}(L_k L_k^{\dagger}),
 \end{equation}

\noindent where in the second approximation we further assumed large number of qubits. As we show in Appendix~\ref{subapp:avgoverlapmarkovian}, within these approximations the coefficient $r_{\mathcal{N}}$ also controls the decay of the average purity, which in turn can be approximated by substituting the associated Haar average channel with coefficient $p_{\mathrm{eff}}$ given in Eq.~\eqref{eq:peffmaintext}. This means that approximating a Lindbladian noise map to first order and for large number of qubits
 
 \begin{equation}
     r_{\mathcal{N}} \approx 1 - 2 p_{\mathrm{eff}}.
 \end{equation}
 
 This suggests that, in this limit, the decay of the variance can be explained by the fact that when we sample random parameters, the system approaches with high probability a fixed state, which for weak noise turns out to be the completely mixed state. We discuss this point in more detail in the next subsection.

\subsubsection{Purity decay}
\label{subsec:puritydecay}

If under the repeated action of the noise, a parameterized quantum state approaches a fixed point $\sigma$ for any parameter, we expect any cost function to concentrate, and thus its gradient to go to zero. This can be shown mathematically by considering the absolute value of the difference between the cost function evaluated on the PQC and the cost function evaluated on the fixed point $C_{\sigma} = \mathrm{Tr}(O \sigma)$:

\begin{equation}
    | C(\bm{\theta}) - C_{\sigma} |
    = |\mathrm{Tr} [ O ( \rho(\bm{\theta})
    -   \sigma ) ]|.
\end{equation}

To bound this expression, similar to the derivation of Eq.~\eqref{eq:nibpparshift}, we exploit the properties of Schatten $p$-norms, which are introduced in Appendix~\ref{app:schatten}. We can write

\begin{equation}
\label{eq:costfrelentropy}
    | C(\bm{\theta}) - C_{\sigma} | 
    \leq ||O||_{\infty} ||   \rho(\bm{\theta}) 
    -   \sigma   ||_{1}  
     \leq  ||O||_{\infty} \sqrt{ 2 \ln(2) D( \rho (\bm{\theta}) || \sigma  )  }.
\end{equation}

\noindent where $D( \rho (\bm{\theta})||\sigma)$ is the quantum relative entropy between $ \rho(\bm{\theta})$ and $\sigma$. In the first line of Eq.~\eqref{eq:costfrelentropy} we used Hölders inequality, given in Eq.~\eqref{eq:holders}, and in the second line we used Pinsker's inequality,  given in Eq.~\eqref{eq:pinskers} in the Appendix. The connection to the relative entropy allows us to connect the concentration of the cost function to the decay of the purity. As shown in Appendix~\ref{app:var_grad_decay}, in the case of a weak Lindbladian noise map, the fixed point is $\sigma \approx I/2^n$, and therefore the relative entropy is bounded by the purity $\mathrm{Tr}(\rho^2)$ \cite{eisert}
 
\begin{equation}
\label{eq:relentropypuriy}
    D\biggl(\rho \biggl \lVert \frac{I}{ 2^n } \biggr) 
    \leq n + \log_2[\mathrm{Tr}(\rho^2)].
\end{equation}

This means that as the purity of $\rho(\bm{\theta})$ approaches the purity of the completely mixed state, the distance between the corresponding cost functions decreases

\begin{equation}
    | C(\bm{\theta}) - C_{I/2^n} | 
    \leq  ||O||_{\infty} \sqrt{ 2 \ln(2) \Bigl \{ n + \log_2[\mathrm{Tr}(\rho(\bm{\theta})^2)] \Bigr \}  }.
\end{equation}

 In what follows, we will be interested in how the purity of a quantum state in our toy model approaches, in the small noise limit, the purity of the completely mixed state, that is $ 1/2^n$.  

Let us start by considering the behavior of the purity on average. As we show in Appendix~\ref{sec:app_c}, this purity can be computed exactly (see Eq.~\eqref{eq:avgoverlap2_all}). We note that the same result has also been obtained in Ref.~\cite{eisert}. In the limit of weak noise, the average purity decrease due to one layer of unitary $2$-design followed by noise is approximately given by the purity decrease due to the Haar averaged channel, as shown in Eq.~\eqref{eq:haareqalavg}:

\begin{equation}
 \int d \mu (U) \mathrm{Tr}[(\mathcal{N} \circ \mathcal{U} (\rho_{\mathrm{in}}))^2 ]
    \approx \mathrm{Tr}[(\mathcal{N}_{\mathrm{Haar}}(\rho_{\mathrm{in}}))^2 ]. 
\end{equation}

\begin{figure}
    \centering
    \includegraphics[width=0.6\textwidth]{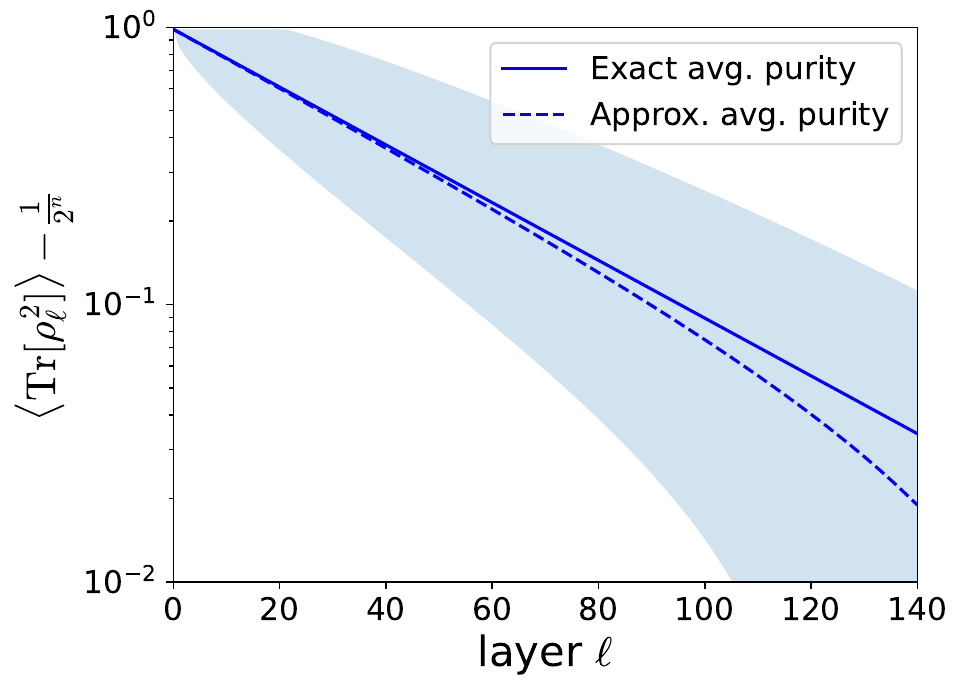}
    \caption{Approximate (Eq.~\eqref{eq:pur_approx}) and exact (Eq.~\eqref{eq:avgoverlap2_all} in Appendix~\ref{sec:app_c}) average purity as a function of the number of layers of the toy model in Fig.~\ref{fig:toy_model} for $n=6$ qubits and single-qubit amplitude damping noise with parameter $\gamma_{\downarrow}=0.004$. The unitaries at each layer are randomly drawn from a unitary $2$-design. The initial state is assumed to be a pure state, i.e., $\mathrm{Tr}(\rho_{\mathrm{in}}^2)=1$. The blue region contains (approximately) more than 99 \% of the single instance purities, according to the bounds obtained in Eq.~\eqref{eq:purityboundpmax} in Appendix~\ref{app:pur_single}.
     }
    \label{fig:hoeffd_bounds}
\end{figure}

 This means that we can approximately obtain the average purity of the state $\rho_{\ell}$ after $\ell$ layers by applying the Haar averaged channel $\ell$ times

\begin{equation}
 \langle \mathrm{Tr}(\rho_{\ell}^2 ) \rangle =
 \int  d \mu (U_1) \dots \int  d\mu (U_\ell) \mathrm{Tr}[(\mathcal{N} \circ \mathcal{U}_{\ell} \circ... \circ \mathcal{N} \circ \mathcal{U}_1 (\rho_{\mathrm{in}}) )^2 )
 \approx  
    \mathrm{Tr}[(\mathcal{N}^{\ell}_{\mathrm{Haar}}(\rho_{\mathrm{in}}))^2 ]. 
\end{equation}

We now assume that for all layers $\ell' \le \ell$ the purity after the application of the Haar averaged channel is much larger than the purity of the completely mixed state, i.e., $\mathrm{Tr}[(\mathcal{N}_{\mathrm{Haar}}(\rho_{\ell'}))^2] \gg 1/2^n$. With this assumption, we can neglect terms that scale as $1/2^n$, and obtain that

\begin{equation}
 \mathrm{Tr}[(\mathcal{N}_{\mathrm{Haar}}(\rho_{\ell'-1}))^2]   
 \approx (1-2 p_{\mathrm{eff}}) \mathrm{Tr}(\rho_{\ell'}^2)  \label{eq:haar_one_layer},
\end{equation}

\noindent $\forall \, \ell' \le \ell$. This leads to a simple exponential decay of the average purity with the number of layers in our model:

\begin{equation}
\label{eq:pur_approx}
\langle \mathrm{Tr}(\rho_{\ell}^2) \rangle \approx (1-2 p_{\mathrm{eff}})^{\ell} \mathrm{Tr}(\rho_{\mathrm{in}}^2). 
\end{equation}

This exponential behavior holds as long as the average purity is not too close to that of the completely mixed state, which approaches zero for large number of qubits. More generally, as shown in Appendix~\ref{sec:app_c} and also obtained in Ref.~\cite{eisert}, the average purity shows an exponential decay towards a fixed value that depends on the noise map $\mathcal{N}$. 

In order to explain the concentration of the cost function, we also need to argue that the purities of the single instances do not deviate too much from the average purity. In Appendix~\ref{app:pur_single}, using Hoeffding's inequality \cite{hoeffding1963}, we obtain bounds on the deviation of the single instances from the average behavior in the weak noise limit. Fig.~\ref{fig:hoeffd_bounds} shows an example of the average purity decay for single-qubit amplitude damping noise characterized by the PTM given in Eq.~\eqref{eq:ptmad}. 

\begin{figure}
    \centering
    \includegraphics[width=\textwidth]{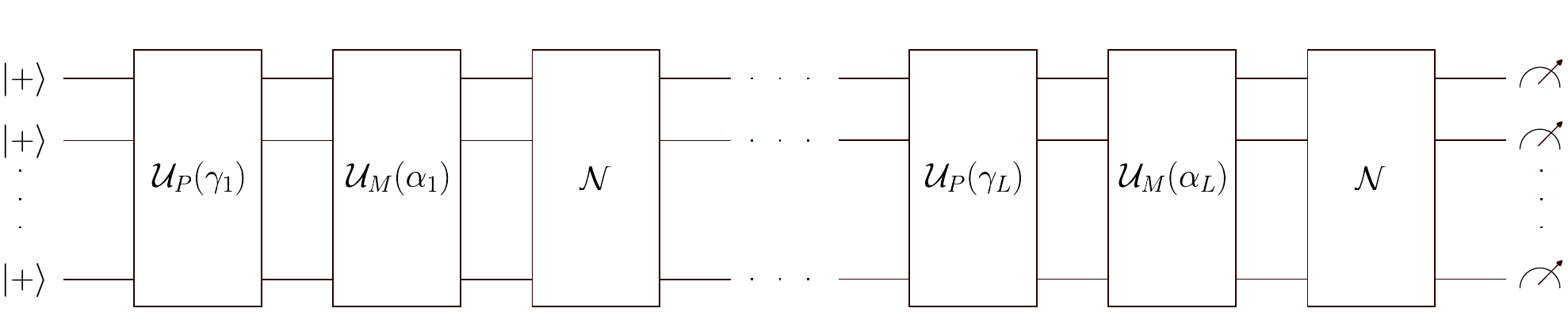}
    \caption{Layered noise model for QAOA. The noise map $\mathcal{N}$ is applied after each layer of QAOA consisting of an application of the channel corresponding to the problem unitary $\mathcal{U}_P(\gamma_{\ell})$ followed by the channel corresponding to the mixer unitary $\mathcal{U}_M(\alpha_{\ell})$ for $\ell =1, \dots, L$.}
    \label{fig:QAOA_noise}
\end{figure}

\section{Numerical results for QAOA applied to MaxCut problems}
\label{sec:numerical}

In this section, we study numerically NIBPs for the case of standard QAOA \cite{farhi2014} focusing on the case of single-qubit amplitude damping noise. In fact, for standard QAOA a parameter-shift rule does not hold and the QAOA unitaries do not satisfy the unitary $2$-design property in general. Thus, QAOA could potentially avoid the analytical results obtained in Sec.~\ref{sec:analytical}. 

We consider QAOA circuits for MaxCut problems on $d$-regular graphs, where each vertex is connected to $d$ other vertices. QAOA performances for these kinds of problems have been extensively studied in the literature \cite{zhou2020, basso2021}. Examples of $d$-regular graphs, that we will use in our numerical analysis, are shown in Fig.~\ref{fig:used_graphs} in the Appendix. 

Given a graph $G=(V, E)$ with $V$ the set of vertices and $E$ the set of edges, QAOA associates a qubit to each vertex and thus $n=|V|$. QAOA consists in the alternated application of parameterized mixing and problem unitaries. In the standard formulation of QAOA the mixing unitary $U_M(\alpha)$ with parameter $\alpha$ is associated with a mixing Hamiltonian $H_M$:

\begin{equation}
    H_M = \sum_{i \in V} X_i, \quad U_M (\alpha) = e^{-i \alpha H_M}.
\end{equation}

For MaxCut the problem Hamiltonian $H_P$ and the corresponding problem unitary $U_P (\gamma)$ with parameter $\gamma$ read

\begin{equation}
    H_P =  \sum_{(i, j)\in E} Z_i Z_j, \quad U_P(\gamma) = e^{-i \gamma H_P}.
\end{equation}

Thus, we define the QAOA unitary up to $L$ layers as

\begin{equation}
    U(\bm{\alpha}, \bm{\gamma}) = 
    e^{- i \alpha_{L} H_{M}}
    e^{- i \gamma_{L} H_{P}} \dots
    e^{- i \alpha_{1} H_{M}}
    e^{- i \gamma_{1} H_{P}},
\end{equation}

\noindent where $\bm{\alpha} = (\alpha_{1}, \dots, \alpha_{L} )$ and $\bm{\gamma} =(\gamma_{1}, \dots,\gamma_{L})$ are vectors with the parameters of the mixing and problem unitaries, respectively.

The ground state of $H_{P}$ corresponds to the solution of the MaxCut problem. In QAOA the quantum system is initialized in the state $\ket{+}^{\otimes n}$ and after $L$ layers the ideal, noiseless quantum state is

\begin{equation}
    \ket{\psi(\bm{\alpha},\bm{\gamma})}
    = U(\bm{\alpha}, \bm{\gamma})  
    \ket{+}^{\otimes n}.
\end{equation}

 The goal of QAOA is to find optimal parameters such that the parameterized quantum state $\ket{\psi(\bm{\alpha}, \bm{\bm{\gamma}})}$ minimizes the average of the problem Hamiltonian $H_P$. Thus, the cost function of QAOA is given by 

\begin{equation}
    C (\bm{\alpha}, \bm{\gamma}) = \mathrm{Tr}[H_P \ket{\psi(\bm{\alpha},\bm{\gamma})} \bra{\psi(\bm{\alpha},\bm{\gamma})}].
\end{equation}

Note that the QAOA circuit can be seen as a discretized version of quantum annealing \cite{kadowaki1998, zhou2020}. 

In what follows, we study noisy QAOA circuits using a layered noise model, as shown in Fig.~\ref{fig:QAOA_noise}. In our numerical analysis, we take the noise map to be single-qubit amplitude damping noise acting independently on each qubit, described in detail in Appendix~\ref{app:subsubsecad}. We will be mostly interested in comparing the numerical results with the results predicted by the toy model where the unitaries form a unitary
$2$-design, that we analyzed in Subsec.~\ref{subsec:nibp2design}.

\begin{figure}
    \centering
    \includegraphics[width=\textwidth]{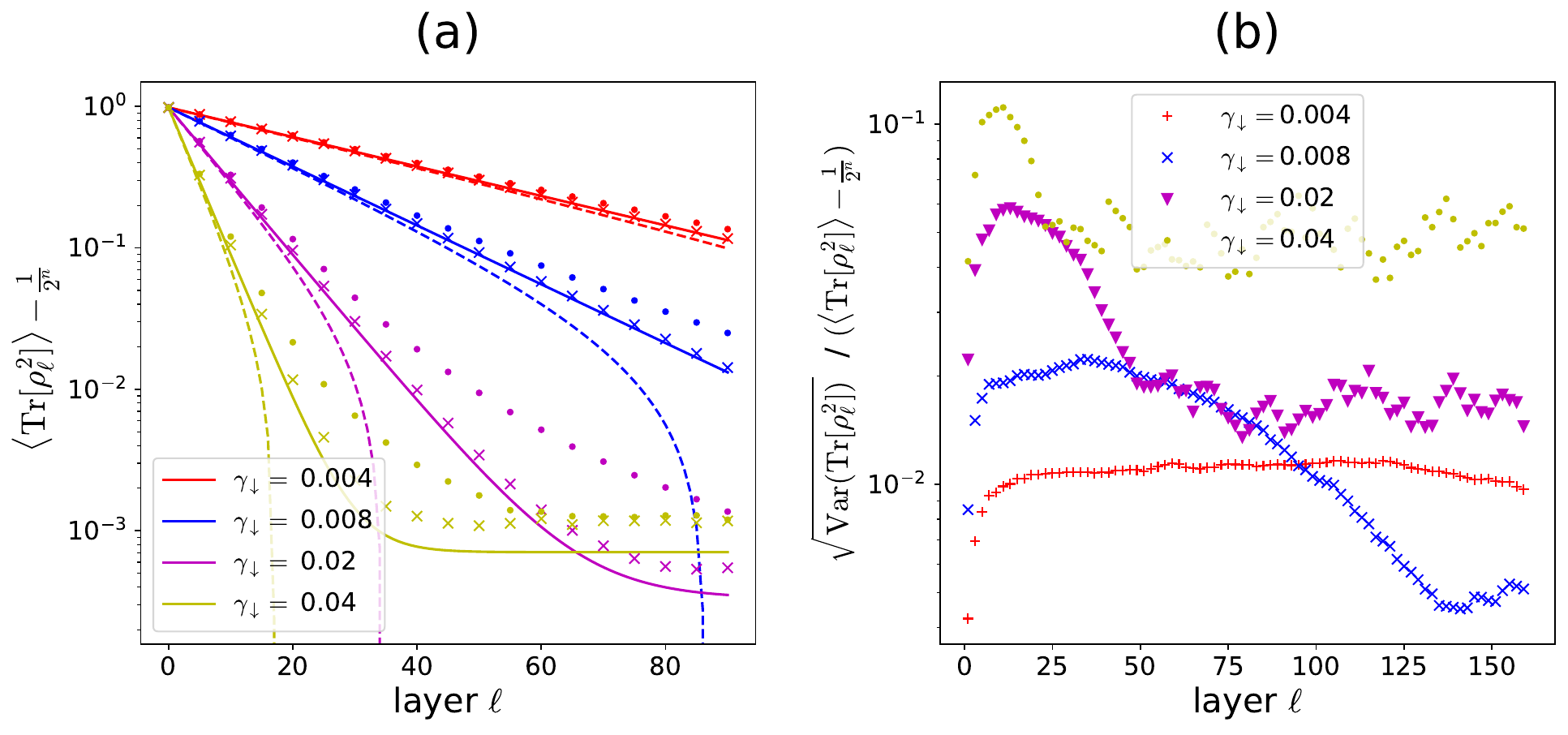}
    \caption{(a) Comparison of the decay of the average purity in QAOA circuits for MaxCut problems and the toy model for different amplitude damping parameters $\gamma_{\downarrow}$. The QAOA results correspond to a $6$-vertex $3$-regular (cross) and a $5$-regular (dots) graph. The solid lines give the exact average purities in the toy model circuit (Eq.~\eqref{eq:avgoverlap2_all} in Appendix~\ref{sec:app_c}). The dashed lines show the approximate purities in the toy model circuit (Eq.~\eqref{eq:pur_approx}). The markers give the numerical average purities in the QAOA circuits for the corresponding noise strength and graph. The averages are evaluated by sampling $128$ times parameters from the uniform distribution $\alpha_{\ell},\gamma_{\ell} \in [0,2 \pi )$. (b) Variance of the purities in the QAOA circuit corresponding to the $6$-vertex $3$-regular graph for different amplitude damping parameters $\gamma_{\downarrow}$.}
    \label{fig:pur_qaoa}
\end{figure}

\subsection{Purity decay in QAOA circuits} \label{subsec:pur_QAOA}

In this section, we compare the average purity in the toy model with the average purity in the QAOA circuits for MaxCut problems. We show the resulting average purity for several noise strengths and the $6$-vertex $3$-regular and $5$-regular graphs in Fig.~\ref{fig:pur_qaoa}(a). We see that the numerical results for QAOA agree well with the analytical ones of the toy model as long as the purity is not too close to that of the completely mixed state. 

Note that for all noise instances, the purity in the circuit corresponding to the $3$-regular graph is described better by the purity in the toy model circuit than the purity corresponding to the $5$-regular graph. In Appendix \ref{appsubsec:twirlingqaoa}, we discuss this further by showing that in the limit of many layers QAOA based on the $3$-regular graph approximates better the behavior of a Haar random channel compared to the $5$-regular graph. However, even for the $5$-regular graph, we find a good match with the toy model, as long as the purity is not too small. 

Besides, we find that in the limit of deep circuits, the exact average purity of the toy model and the average purities from the QAOA circuits do not go to the purity corresponding to the completely mixed state but reach a plateau at slightly larger values. We attribute this to the fact that the completely mixed state is not a fixed point under amplitude damping. Thus, if the system reaches the completely mixed state, amplitude damping can increase the purity of the state again and the average purity reaches a plateau.

Fig.~\ref{fig:pur_qaoa}(b) shows the behavior of the variance of the purity for the case of the $3$-regular graph. We see that with increasing circuit depths the variance first increases and then decreases again. The reason for the increase is that as noise enters the system, the purities of the single instances can deviate more from the average. 

However, as the number of layers increases, most instances get close to the completely mixed state which makes the variance decrease again. Furthermore, similar to the average purity the variance exhibits plateaus at a certain minimal value. These plateaus correspond to fluctuations around the completely mixed state. Larger noise parameters $\gamma_{\downarrow}$ correspond to larger fluctuations.  

In App.~\ref{appsubsec:more_purs_qaoa} we give additional numerical data for QAOA circuits with larger qubit number.

\subsection{Decay of the gradient of the cost function} \label{sec:grad}

\begin{figure}
\centering
    \includegraphics[width=0.6\textwidth]{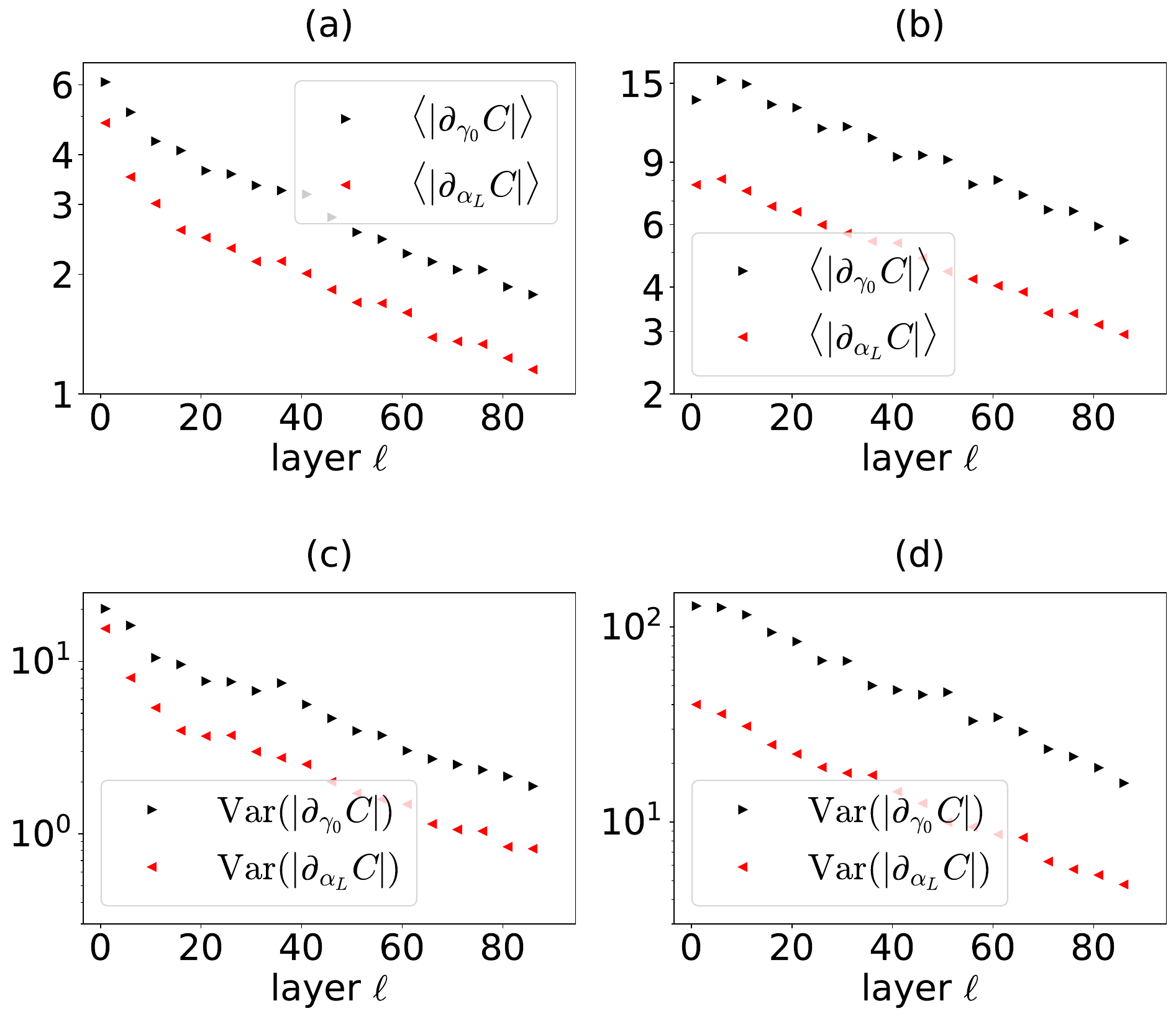}
    \caption{    Average and variance of the partial derivative with respect to the first ($\gamma_0$) and last ($\alpha_L$) parameter in QAOA for $6$-vertex graphs for a $3$-regular graph ((a) and (c)) and a $5$-regular graph ((b) and (d)). The amplitude damping parameter $\gamma_{\downarrow}$ is fixed to $\gamma_{\downarrow}= 0.004$. Averages and variances are evaluated by taking $640$ samples with parameters chosen randomly from the uniform distribution $\alpha_{\ell},\gamma_{\ell} \in [0,2 \pi )$.}
    \label{fig:derivatives}
\end{figure}

As the purity of the quantum state approaches the purity of the completely mixed state, we also expect that the gradient of the cost function decreases. In the case of weak Lindbladian noise this is the effect causing the emergence of NIBPs as discussed in Subsec.~\ref{subsec:nibp2design}. 

To study this phenomenon in our QAOA circuits, we compute numerically the derivative of the cost function with respect to the first and the last circuit parameters. In particular, we consider $\partial_{\gamma_{0}} C:= \partial C / \partial \gamma_{0}$ and $\partial_{\alpha_{L}} C:= \partial C / \partial \alpha_{L}$. 

 We consider the average and the variance of the absolute values of the derivatives in Fig.~\ref{fig:derivatives} for the circuits corresponding to the $3$- and $5$-regular graphs in the presence of amplitude damping noise. We see that both the average and the variance are decreasing exponentially with the number of layers. Note that in the figure the exponential is mild because we take a small noise parameter $\gamma_{\downarrow}=0.004$ corresponding to the red curves in the purity plots shown in Fig.~\ref{fig:pur_qaoa}. This shows the existence of NIBP in the QAOA circuits we considered.

\section{Conclusion} 
\label{sec:conclusion}

In this work, we studied the emergence of NIBPs in VQAs under general layered noise models. While a recent contribution proved the existence of NIBPs in variational quantum circuits under local Pauli noise~\cite{Wang}, the picture for more general noise maps is less clear. To approach this issue, we considered two cases that both allow for an analytical treatment. In the first case, we discussed NIBPs in PQCs for which the partial derivatives of the cost function can be evaluated via a parameter-shift rule. We found that in this case, NIBPs emerge as a simple consequence of the fact that CPTP maps are contractive with respect to the trace distance. In the second case, we considered a toy model circuit that consists of alternating applications of unitary $2$-designs and noise. The unitary $2$-design property captures the fact that the variational ansatz is sufficiently expressive of the full unitary group. Highly expressive unitaries have been linked to the emergence of SBPs \cite{McClean2018, holmes2022}, and have  been used to study the limitations of error mitigation techniques \cite{eisert}. Our work shows that highly expressive unitaries that approximate the $2$-design property will also give rise to NIBPs even if a parameter-shift rule does not hold. Furthermore, for weak Linbladian noise, we show that the decay of the variance of the gradient is linked to the decay of the average purity in the model.

Our analytical results suggest that in order to mitigate the NIBPs problem, PQCs for which a parameter-shift rule holds, as well as highly expressive PQCs, should be avoided. The second requirement is also needed to avoid SBPs \cite{holmes2022} and hints at the fact that the variational ansatz should be tailored to the problem, in order to be successful. QAOA circuits seem to match these conditions and formally escape our analytical derivations. Thus, we studied numerically the emergence of NIBPs in QAOA circuits, focusing on the case of MaxCut problems for small $d$-regular graphs and amplitude damping noise. We found that the analytical predictions of the toy model with unitary $2$-designs still provide a good approximation of the behavior of the purity in QAOA circuits, especially for weak noise and for more expressive graphs. Additionally, our numerics show that NIBPs are emerging for the circuits we considered and amplitude damping noise. While our results are problem-specific, they suggest that NIBPs seem to emerge in typical QAOA circuits subject to amplitude damping noise. 

\subsection{Outlook}

An open question for QAOA circuits is to understand whether given a problem instance, a noise model and a noise strength, the required QAOA depth to solve the problem to a desired accuracy is small enough that NIBPs would not be a \emph{practical} issue in the optimization process. Apart from the barren plateau problem, Ref.~\cite{cirac2022} (see also the analysis in Ref.~\cite{Lotshaw2022}) has estimated, with yet a different error model, that QAOA circuits can become useful only at error rates that are few orders of magnitude below those of fault-tolerant thresholds.

The layered error model we analyzed is a Markovian error model, where the state after each layer only depends on the state at the layer before. A recent work \cite{burkard2023} has analyzed the impact of correlated, non-Markovian noise in QAOA circuits finding that this kind of noise can be beneficial for the performance of the algorithm. It is an interesting question to understand if these kinds of noise models can tame the NIBP problem in VQAs. Furthermore, it would be insightful to perform random QAOA experiments as that we numerically simulated in Fig.~\ref{fig:pur_qaoa} on real quantum hardware where several, different noise sources are present. To perform these kinds of experiments one simply samples randomly the QAOA parameters for the desired depth and obtains the purity of the quantum state on the output. Possible ways of measuring the purity of the quantum state on NISQ devices based on randomized measurements are discussed in Ref.~\cite{Elben2023}.

During the writing of this manuscript we became aware of two recent publications, namely Refs.~\cite{fontana2023, ragone2023}, that connect different sources of the barren plateau phenomenon to Lie algebraic properties of the PQC. In particular, it would be interesting to see whether these properties can also be used to relax the unitary 2-design property in our toy model. We also highlight two recent, independent and complementary approaches for the study of NIBPs beyond unital noise that have been put forward in Refs.~\cite{singkanipa2024unital, mele2024}.

\section*{Data availability statement}

The data that support the findings of this study are openly available at the following URL/DOI: https://github.com/maschuu/data\_QST

\section*{Acknowledgements}
We thank Daniel Stilck França for useful comments on the manuscript. A.C. acknowledges funding from the Deutsche Forschungsgemeinschaft (DFG, German Research Foundation) under Germany's Excellence Strategy – Cluster of Excellence Matter and Light for Quantum Computing (ML4Q) EXC 2004/1 – 390534769. A.C. and M.S. acknowledge funding from the German Federal Ministry of Education and Research (BMBF) in the funding program ``Quantum technologies – from basic research to market" (Project QAI2, contract number 13N15585). M. S. was also funded by the BMBF in the funding program ``Quantum technologies – from basic research to
market" (Project QSolid, contract number 13N16149).

\section*{Conflict of interest}

The authors declare no competing interest.

\vspace{2cm}

\bibliography{biblio_nibp}

\newpage

\appendix 

\numberwithin{equation}{section}

\section{Preliminary concepts and notation}
\label{app:preliminary}
We consider a system of $n$ qubits. The Hilbert space of a single-qubit is $\mathcal{H} = \mathbb{C}^2$. The Hilbert space for $n$ qubits $\mathcal{H}_n$ is immediately obtained introducing a tensor product structure $\mathcal{H}_n = \mathcal{H}^{\otimes n} = \mathbb{C}^{2n}$. Only rays in the Hilbert space correspond to physical states and so we can always take a pure quantum state to be a unit vector $\ket{\psi} \in \mathcal{H}_n$. Let $\mathcal{B}_n$ be the vector space of linear bounded operators in $\mathcal{H}_n$. A generic mixed density matrix $\rho$ is a positive semidefinite operator in $\mathcal{B}_n$ with unit trace. We denote by $\mathrm{D}_n$ the set of density matrices $\mathrm{D}_n \subset \mathcal{B}_n$. General quantum operations are completely positive, trace-preserving (CPTP) maps $\mathcal{E}$ that map a density matrix to another density matrix, i.e., $\mathcal{E}(\rho) = \rho' \in \mathrm{D}_n, \quad \forall \rho \in \mathrm{D}_n$. We define the single-qubit Pauli operators in the computational basis as the following $2 \times 2$ matrices:

\begin{equation}
\begin{aligned}
I = \begin{pmatrix}
1 & 0 \\
0 & 1
\end{pmatrix},& \quad X = \begin{pmatrix}
0 & 1 \\
1 & 0
\end{pmatrix},
 \\
Y = \begin{pmatrix}
0 & -i \\
i & 0
\end{pmatrix},& \quad Z = \begin{pmatrix}
1 & 0 \\
0 & -1
\end{pmatrix}.
\end{aligned}
\end{equation} 
We define the Pauli operators for $n$ qubits as \footnote{$\mathbb{F}_2$ is the binary field, i.e., $\mathbb{F}_2 = \{0, 1 \}$, so $\mathbb{F}_{2}^{2 n}$ is a $2n$-dimensional binary vector.}

\begin{equation}
\label{eq:pw}
\pw{r} = i^{x^T z}X(x) Z(z)  ,\quad r = \begin{pmatrix}
x \\ z
\end{pmatrix} \in \mathbb{F}_2^{2 n},
\end{equation}
where we denoted compactly the operators

\begin{subequations}
\begin{align}
X(x) &= X^{x_1} \otimes \dots \otimes X^{x_n}, \\
Z(z) &= Z^{z_1} \otimes \dots \otimes Z^{z_n},
\end{align}
\end{subequations}
with

\begin{subequations}
\begin{align} 
x &= \begin{pmatrix}
x_1 & \dots & x_n
\end{pmatrix}^T \in \mathbb{F}_2^n, \\ 
z &= \begin{pmatrix}
z_1 & \dots & z_n
\end{pmatrix}^T \in \mathbb{F}_2^n.
\end{align} 
\end{subequations}
Explicitly with this parameterization the single-qubit Pauli operators are $\pw{0, 0} = I, \pw{0, 1} = Z, \pw{1, 0} = X, \pw{1, 1} = Y$. Notice that according to our definition they are unitary, traceless and hermitian, \i.e., $\pw{r} = \pw{r}^{\dagger}$. This is however our arbitrary choice and the phase factor $i^{x^T z}$ could also be chosen as $e^{i \pi x^T z}$ in which case we would get $i Y$ in place of $Y$. The Pauli group for $n$ qubits is defined as 

\begin{equation}
\mathcal{P}_n = \{i^{\kappa} \pw{r} \lvert r \in \mathbb{F}_2^{2 n}, \, \kappa \in \{0, 1, 2, 3 \} \}.
\end{equation}
The order of $\mathcal{P}_n$, \i.e., the number of elements in $\mathcal{P}_n$, is $\lvert \mathcal{P}_n \rvert = 4^{n +1}$. The Pauli operators in Eq.~\eqref{eq:pw} are the elements of the Pauli group with positive phase, that is with $\kappa = 0$. We denote the normalized Pauli operators by $\pwn{r}$ that are related to the $\pw{r}$ simply by a proportionality factor

\begin{equation}
\pwn{r} = \frac{1}{2^{n/2}} \pw{r}.
\end{equation}
We can expand any operator $A \in \mathcal{B}_n$ as a linear combination of Pauli operators

\begin{equation}
A = \sum_{r \in \mathbb{F}_2^{2n}} \chi_{A}(r) \pwn{r},
\end{equation}
where $\chi_{A}(r)$ is called the characteristic function of $A$ and it is given by

\begin{equation}
\chi_{A}(r) = \mathrm{Tr}(\pwn{r} A).
\end{equation}
Thus, we can ``vectorize" the operator $A$ and write $\kket{A}$ to denote the $4^n$-dimensional column vector with $\chi_A(r)$ as its entries. We denote by $\bbra{A}$ the conjugate transpose of $\kket{A}$ in complete analogy with the standard treatment of pure states in quantum mechanics. We have effectively made $\mathcal{B}_n$ a Hilbert space with a inner product

\begin{equation}
\label{eq:hs}
\bbrakket{A}{B} = \mathrm{Tr}(A^{\dagger} B),
\end{equation} 
that is usually called Hilbert-Schmidt inner product. The normalized Pauli matrices are orthonormal with respect to this inner product since

\begin{equation}
\bbrakket{\pwn{r}}{\pwn{r'}} = \delta_{r r'}.
\end{equation}
Again in complete analogy with the standard pure state formalism of quantum mechanics we get that

\begin{equation}
\sum_{r \in \mathbb{F}_2^{2n}} \kket{\pwn{r}} \bbra{\pwn{r}} = \hat{\mathcal{I}},
\end{equation}
where $\hat{\mathcal{I}}$ is the $4^n \times 4^n$ identity matrix. Notice that this treatment is not only limited to the Pauli basis, but we can do the same for any Hilbert-Schmidt basis for $\mathcal{B}_n$. For instance, if $U$ is a generic unitary then $U \pwn{r} U^{\dagger} $ is again an orthonormal basis for $\mathcal{B}_n$. In what follows we will always assume that $\mathcal{B}_n$ is a Hilbert space equipped with the Hilbert-Schmidt inner product Eq.~\eqref{eq:hs}. We will refer to the formalism where we use vectorized operators $\kket{A}$ as the Liouville representation of quantum mechanics. 

In the Liouville representation, the action of a linear map $\mathcal{M}$, not necessarily CPTP, on an operator matrix $A$ can be expressed simply as a $4^{n} \times 4^{n}$ matrix that we denote by $\hat{\mathcal{M}}$ acting on $\kket{A}$. In the normalized Pauli basis the elements of this matrix are given by

\begin{equation}
\label{eq:ptmel}
\hat{\mathcal{M}}_{r r'} = \bbra{\pwn{r}} \hat{\mathcal{M}} \kket{\pwn{r'}} = \mathrm{Tr}[\pwn{r} \mathcal{M}(\pwn{r'} )],
\end{equation} 
which is called the Pauli Transfer Matrix (PTM). If $\mathcal{M}$ is a CPTP map then it would act on vectorized density matrices as

\begin{equation}
\kket{\rho'} = \hat{\mathcal{M}} \kket{\rho} \in \mathrm{D}_n, \quad \forall \kket{\rho} \in \mathrm{D}_n.
\end{equation}

\subsection{Basics of representation theory}
\label{subapp:basicgroup}

We begin by reviewing the statement of Schur's lemma \cite{zee}. We denote by $R(g)$ a representation of a group $G$ with $g$ a group element $g \in G$. We say that a representation $R(g)$ of a group $G$ is $d$-dimensional if $R(g)$ is a $d \times d$ complex matrix, i.e., $R(g) \in \mathbb{C}^d \times \mathbb{C}^d,\, \forall g \in G.$   Representations of finite groups can always be taken to be unitary. It is immediate to realize that given a $d$-dimensional representation $R(g)$ we can construct a $2d$-dimensional representation $R'(g)$ simply by copying $R(g)$ twice as

\begin{equation}
R'(g) = \begin{pmatrix}
R(g) & 0 \\ 
0 & R(g)
\end{pmatrix} =  \bigoplus_{k=1}^2 R(g)
\end{equation}
We can also repeat the previous trick $m$ times and obtain a $dm$-dimensional representation $R'(g) = \bigoplus_{k=1}^m R(g)$. More generally, if we have a set of $m$ representations $R_1(g), \dots, R_{m}(g)$ with dimensions $d_1, \dots, d_m$, respectively, we can construct a new one with dimension $d= \sum_{k=1}^m d_k$, as 

\begin{equation}
\label{eq:block_rep}
R'(g) = \bigoplus_{k=1}^m R_{k}(g).
\end{equation}
We say that a representation as in Eq.~\eqref{eq:block_rep} is block-diagonal or equivalently that is a direct sum of smaller representations. Notice that it is possible that the $R_{k}(g)$ could in turn be written as a direct sum of smaller representations. Furthermore, we also remark that if we write $R'(g)$ in a different basis we obtain a representation $R''(g) =S^{-1} R'(g) S$ with $S$ the matrix representing the change of basis. We consider $R''(g)$ as the same representation as $R'(g)$ by saying that $R''(g)$ and $R'(g)$ are related by a similarity transformation. However, while $R'(g)$ is block-diagonal,  $R''(g)$ is generally not and so we would need to figure out what $S$ is in order to bring it into a manifestly block-diagonal form. 

\begin{defn}
A representation $R(g)$ of a group $G$ is said to be irreducible if it cannot be written as a direct sum of smaller representations as in Eq.~\eqref{eq:block_rep}.
\end{defn}

We will refer to the irreducible representations of a group $G$ as the irreps. of $G$. The importance of the characterization of the irreps. of a group $G$ stems from Maschke's theorem (or complete reducibility) that states that all representations of $G$ can be written, after suited similarity transformations, as direct sums of irreps \cite{fulton_harris}. 

\begin{lem}
\label{lem:schur}
\emph{(Schur's lemma)} If $R(g)$ is an irreducible representation of a finite or compact group $G$ and there is a matrix $A$ that commutes with $R(g) \, \forall g \in G$, i.e., $A R(g) = R(g) A$, then $A = \lambda I$, with $\lambda \in \mathbb{C}$ and $I$ the identity.
\end{lem}

We refer the reader to Ref.~\cite{zee} for the proof of Schur's lemma, but it can also be found in most standard references on group theory. In what follows, we will need also Schur's lemma for reducible representations, but where the same representation is not repeated, i.e., with no multiplicities. 

\begin{lem}
\label{lem:schur2}
\emph{(Schur's lemma for reducible representations with no multiplicities)} If $R(g)$ is a reducible representation of a finite or compact group $G$ with no multiplicity, that is $R(g) = \bigoplus_{k=1}^m R_k(g) = \sum_{k=1}^m R_{k}(g) \Pi_k$ with $\Pi_k$ orthogonal projectors and there is a matrix $A$ that commutes with $R(g) \, \forall g \in G$, i.e., $A R(g) = R(g) A$, then $A = \sum_{k=1}^{m} \lambda_k \Pi_k$, with $\lambda_k \in \mathbb{C}$.
\end{lem}

This result can be obtained by applying the standard Schur's lemma, Lemma~\ref{lem:schur}, to each block of the matrix.

\subsection{Unitary $1$- and $2$-designs} \label{app:unitary_2}
We denote by $d \mu(U)$ the Haar random measure on the unitary group $\mathrm{U}(2^n)$ of $2^n \times 2^n$ unitary matrices. Notice that $\mathrm{U}(2^n)$ is a compact group and so Schur's lemma can be applied. Integrals over $d \mu(U)$ have to be intended as integrals over the whole unitary group $\mathrm{U}(2^n)$. Let $A$ be an arbitrary operator acting on $n$ qubits and consider the operator

\begin{equation}
A_{\mathrm{Haar}} = \int d\mu (U) U^{\dagger} A U.
\end{equation} 
This operator commutes with every $V \in \mathrm{U}(2^n) $ and thus it satisfies the assumptions of Schur's lemma. In addition, $U$ is the defining representation of $\mathrm{U}(2^n)$. Thus, we conclude that $A_{\mathrm{Haar}}$ must be proportional to the identity

\begin{equation}
A_{\mathrm{Haar}} = \lambda I.
\end{equation}
Noticing that $\mathrm{Tr}(A) = \mathrm{Tr}(A_{\mathrm{Haar}})$, we also get $\lambda = \mathrm{Tr}(A)/2^n$. Thus, 

\begin{equation}
\label{eq:haar1}
A_{\mathrm{Haar}} = \int d\mu (U) U^{\dagger} A U = \frac{\mathrm{Tr}(A)}{2^n} I.
\end{equation}
A unitary $1$-design is a set of unitaries $\{U_1, \dots, U_k\}$ \footnote{For simplicity here it is assumed to be discrete, but it can also be continuous}, which can be equipped with a probability distribution over the set $p_k$, such that by averaging over this set an arbitrary operator $A$ we get the same result as the Haar random average in Eq.~\eqref{eq:haar1}. Mathematically,

\begin{equation}
\label{eq:un1design}
\int d\mu (U) U^{\dagger} A U = \sum_k p_k U_k^{\dagger} A U_k = \frac{\mathrm{Tr}(A)}{2^n} I.
\end{equation}
For instance, the Pauli group equipped with the uniform distribution is a $1$-design. 

Let $\hat{\mathcal{U}}$ denote the Liouville representation of a member of the $2^n \times 2^n$ unitary group $\mathrm{U}(2^n)$, i.e., $U \in \mathrm{U}(2^n)$. From the point of view of group theory $\hat{\mathcal{U}}$  is a $4^n$-dimensional representation of $\mathrm{U}(2^n)$. It is immediate to see that this representation breaks into the direct sum of the trivial representation, i.e., 1, and a $(4^n -1) \times (4^n-1)$ representation. In fact, a generic $\hat{\mathcal{U}}$ has the following block-diagonal form

\begin{equation}
\label{eq:ured}
\hat{\mathcal{U}} = \begin{pmatrix} 
1 & \bm{0}_{4^n -1}^T \\
\bm{0}_{4^n -1} & \hat{R}_{U} \end{pmatrix} = 1 \kket{\pwn{0}} \bbra{\pwn{0}} + \hat{R}_{U} \hat{\Pi},
\end{equation}
where $\bm{0}_{4^n -1}$ is a $(4^n-1)$-dimensional column vector of zeros, $\hat{R}_{U}$ is a $(4^n -1)$-dimensional representation and we defined the projector

\begin{equation}
\label{eq:proj}
\hat{\Pi} = \hat{\mathcal{I}} -\kket{\pwn{0}} \bbra{\pwn{0}}.
\end{equation}
$\hat{R}_{U}$ is irreducible, although we do not prove it here. Let us consider an arbitrary linear map $\hat{\mathcal{M}}$, i.e., its Liouville representation. Now, consider the Haar averaged superoperator defined as

\begin{equation}
\hat{\mathcal{M}}_{\mathrm{Haar}} = \int d \mu(U) \hat{\mathcal{U}}^{\dagger} \hat{\mathcal{M}} \hat{\mathcal{U}}. \label{eq_haar2}
\end{equation}
$\hat{\mathcal{M}}_{\mathrm{Haar}}$ commutes with every $\hat{\mathcal{V}}$ with $V \in \mathrm{U}(2^n)$. Since $\mathcal{U}$ can be written as a direct sum of two irreducible representations, i.e., the identity $1$ and $\hat{R}_{U}$ as in Eq.~\eqref{eq:ured}, we can apply Schur's lemma for reducible representations without multiplicities (Lemma~\ref{lem:schur2}) to get that

\begin{equation}
\hat{\mathcal{M}}_{\mathrm{Haar}} = \lambda_0 \kket{\pwn{0}} \bbra{\pwn{0}} + 
\lambda_1 \hat{\Pi}.
\end{equation}
By noticing again that $\mathrm{Tr}(\hat{\mathcal{M}}) = \mathrm{Tr}(\hat{\mathcal{M}}_{\mathrm{Haar}})$, we can determine $\lambda_0$ and $\lambda_1$ to get the final result

\begin{equation}
\label{eq:haar2}
\hat{\mathcal{M}}_{\mathrm{Haar}} = \mathrm{Tr}[\hat{\mathcal{M}} \kket{\pwn{0}} \bbra{\pwn{0}}]\kket{\pwn{0}} \bbra{\pwn{0}}  
+ \frac{\mathrm{Tr}[\hat{\mathcal{M}} \hat{\Pi} ]}{4^n -1} \hat{\Pi}.
\end{equation}
A unitary $2$-design is a set of unitaries $\{U_1, \dots, U_k\}$ that can be equipped with a probability distribution over the set $p_k$, such that by averaging over this set an arbitrary superoperator $\hat{M}$ we get the same result as the Haar random average in  Eq.~\eqref{eq_haar2}. Mathematically

\begin{equation}
\label{eq:haar3}
\int d\mu (U) \hat{\mathcal{U}}^{\dagger} \hat{\mathcal{M}} \hat{\mathcal{U}} = \sum_k p_k \hat{\mathcal{U}}_k^{\dagger} \hat{\mathcal{M}} \hat{\mathcal{U }}_k = 
 \mathrm{Tr}[\hat{\mathcal{M}} \kket{\pwn{0}} \bbra{\pwn{0}}]\kket{\pwn{0}} \bbra{\pwn{0}}  + 
 \frac{\mathrm{Tr}[\hat{\mathcal{M}} \hat{\Pi}]}{4^n -1} \hat{\Pi}.
\end{equation}
A unitary $2$-design is also a unitary $1$-design. To derive this from the previous formulas, consider the superoperator $\mathcal{M}_A$ that acts on an operator $O$ as $\mathcal{M}_A(O) = \mathrm{Tr}(O) A$, with $A$ an arbitrary operator. Using this superoperator in Eq.~\eqref{eq:haar3} gives, after some algebra, the unitary $1$-design property in Eq.~\eqref{eq:un1design} for every $A$.  

When applied to a CPTP map $\mathcal{N}$, Eq.~\eqref{eq:haar2} is simply the usual twirling result in disguise. In fact, considering a vectorized density matrix $\kket{\rho}$, the Haar random averaged channel $\hat{\mathcal{N}}_{\mathrm{Haar}}$ gives

\begin{equation}
\label{eq:twirl1}
\hat{\mathcal{N}}_{\mathrm{Haar}} \kket{\rho} = (1 - p_{\mathrm{eff}}) \kket{\rho} + p_{\mathrm{eff}} \kket{I/2^n}, 
\end{equation}
where 

\begin{equation}
\label{eq:ptwirl}
p_{\mathrm{eff}} = 1 - \frac{\mathrm{Tr}(\hat{\mathcal{N}}) - 1}{4^n-1}.
\end{equation}
In order to obtain Eq.~\eqref{eq:twirl1} we used that for a CPTP map

\begin{equation}
\mathrm{Tr}[\hat{\mathcal{N}} \kket{\pwn{0}} \bbra{\pwn{0}}] = \bbra{\pwn{0}} \hat{\mathcal{N}} \kket{\pwn{0}} = 
\frac{1}{2^{n}} \mathrm{Tr}[\mathcal{N}(I)] = 1.
\end{equation}
In terms of the standard representation of density matrices and quantum operations, Eq.~\eqref{eq:twirl1} becomes

\begin{equation}
\mathcal{N}_{\mathrm{Haar}} (\rho) = \int d \mu(U) \mathcal{U}^\dagger \circ \mathcal{N} \circ \mathcal{U}(\rho) = 
(1 - p_{\mathrm{eff}}) \rho + p_{\mathrm{eff}} \frac{I}{2^n}.
\end{equation}

\subsection{Schatten norms and some of their properties} \label{app:schatten}
We introduce the concept of Schatten $p$-norm and some properties that are used in the main text. The Schatten $p$-norm of an operator $A$ is defined

\begin{equation}
\label{eq:schatten}
    ||A||_{p} = \Bigl( \mathrm{Tr} [ |A|^{p} ]  \Bigr)^{\frac{1}{p}}
\end{equation}
with $|A| = \sqrt{A^{\dagger} A}$. The tracial matrix Hölders inequality states that for two square matrices $A, B$, and $1\leq p, q\leq \infty$ such that $1/p + 1/q =1$, we have

\begin{equation}
\label{eq:holders}
    |\mathrm{Tr}[A^{\dagger} B]|\leq ||A||_{p} ||B||_{q}.
\end{equation}
Besides, Pinsker's inequality gives a lower bound for the relative entropy 

\begin{equation}
   D(\rho||\sigma) = \mathrm{Tr}[\rho \log(\rho)] - \mathrm{Tr} [ \rho \log(\sigma)  ] 
\end{equation}
between two quantum states $\rho$ and $\sigma$:

\begin{equation}
\label{eq:pinskers}
D(\rho||\sigma) \geq \frac{1}{2 \ln(2)} ||\rho- \sigma||_{1}^{2}.
\end{equation}

\subsection{Lindbladian channels}
\label{subapp:markovian}

Lindbladian channels are those channels that can be written as the exponential of a Lindblad dissipator. Here, we focus on Linbladian channels that do not have a coherent, Hamiltonian component. For these channels, the Liouville superoperator can be written as

\begin{equation}
\hat{\mathcal{N}} = e^{ \sum_k \hat{\mathcal{D}}[L_k]},
\end{equation}
where $\hat{\mathcal{D}}[L_k]$ is the superoperator associated with a Lindblad dissipator with jump operators $L_k$ that acts on an operator $A$ as

\begin{equation}
\label{eq:linbladdis}
\mathcal{D}[L_k](A) = L_k A L_k^{\dagger} - \frac{1}{2} L_k^{\dagger} L_k A - \frac{1}{2} A L_{k}^{\dagger} L_k.
\end{equation}
Without loss of generality the jump operators $L_k$ can be taken to be traceless \cite{manzano2020}.

If the noise is small we can approximate a Lindbladian channel to first order as

\begin{equation}
\label{eq:weakmarkoviannoise}
\hat{\mathcal{N}} \approx \hat{\mathcal{I}} + \sum_{k} \hat{\mathcal{D}}[L_k].
\end{equation}
In what follows, it will be useful to express $\mathrm{Tr}(\hat{\mathcal{D}}[L_k])$ in terms of jump operators. In fact, 

\begin{multline}
   \mathrm{Tr}(\hat{\mathcal{D}}[L_k]) = \frac{1}{2^n} \sum_{r \in \mathbb{F}_{2}^{2n}}  \mathrm{Tr}\biggl[  \pw{r} \biggl( L_k \pw{r} L_{k}^{\dagger} - \frac{1}{2} L_{k}^{\dagger} L_{k} \pw{r} - \frac{1}{2} \pw{r} L_{k}^{\dagger} L_k \biggr) \biggr]  \\
   = \frac{1}{2^n} \sum_{r \in \mathbb{F}_{2}^{2n}}  \bigl \{\mathrm{Tr}\bigl[L_{k}^{\dagger}   \pw{r}  L_k \pw{r} \bigr] -  \mathrm{Tr} \bigl[L_{k}^{\dagger} L_k \bigr] \bigr \}.
\end{multline}
Since the Pauli group forms a $1$-design and the jump operators can be taken to be traceless, it follows that

\begin{equation}
    \frac{1}{2^n} \sum_{r \in \mathbb{F}_{2}^{2n}}  \mathrm{Tr}\bigl[L_{k}^{\dagger}   \pw{r}  L_k \pw{r} \bigr] =0, \quad \forall k.
\end{equation}
Thus, 

\begin{equation}
\label{eq:trdjump}
    \mathrm{Tr}(\hat{\mathcal{D}}[L_k]) = - 2^n \mathrm{Tr} \bigl[L_{k}^{\dagger} L_k \bigr] \le 0,
\end{equation}
which follows from the fact that $L_k^{\dagger} L_k$ is a positive operator.

\subsection{Structure of the PTM of relevant superoperators}
\label{app:ptmstructure}
In this subsection, we analyze the structure of the PTM for some relevant superoperators in our analysis. The matrix elements of the PTM of a general superoperator are given in Eq.~\eqref{eq:ptmel}. For a CPTP map $\mathcal{E}$ the PTM has the following form

\begin{equation}
\label{eq:cptpptm}
    \hat{\mathcal{E}} = \begin{pmatrix}
    1 & \bm{0}^T \\
    \bm{s} & \hat{\mathcal{E}}_{\Pi}
    \end{pmatrix},
\end{equation}
where $\bm{0}$ is a $(4^n-1)$-dimensional column vector of zeros, $\bm{s}$ a $(4^n-1)$-dimensional vector and $\hat{\mathcal{E}}_{\Pi}$ is a $(4^n-1)\times(4^n-1)$-dimensional matrix that is the projection of $\hat{\mathcal{E}}$ onto the subspace associated with the projector $\hat{\Pi}$ in Eq.~\eqref{eq:proj}. Consequently, we also have 

\begin{equation}
\label{eq:cptpptmdag}
\hat{\mathcal{E}}^{\dagger} = \begin{pmatrix}
1 & \bm{s}^T \\
\bm{0} & \hat{\mathcal{E}}_{\Pi}^{\dagger}
\end{pmatrix}.
\end{equation}
If the CPTP map is a unitary $\mathcal{U}$ then $\bm{s}$ is also a vector of zeros and so 

\begin{equation}
\hat{\mathcal{U}} = \begin{pmatrix}
1 & \bm{0}^T \\
\bm{0} & \hat{\mathcal{U}}_{\Pi}
\end{pmatrix},
\end{equation}
where $\hat{\mathcal{U}}_{\Pi}$ is in turn a unitary. Notice that if $\mathcal{E}$ is unital, i.e., $\mathcal{E}(I)=I$ then $\bm{s}=0$ as well. 

Let us now consider the dissipative Lindbladian channels that we discussed in Appendix~\ref{subapp:markovian}. Given the action of a Lindblad dissipator described in Eq.~\eqref{eq:linbladdis}, the PTM of a Lindblad dissipator has the form

\begin{equation}
\label{eq:lindbladptm}
    \hat{\mathcal{D}}[L] = \begin{pmatrix}
    0 & \bm{0}^T \\
    \bm{s} & \hat{\mathcal{D}}[L]_{\Pi}
    \end{pmatrix}.
\end{equation}
In fact, one can check that taking the matrix exponential of the matrix on the right-hand side of Eq.~\eqref{eq:lindbladptm} gives a valid PTM of a CPTP map as in Eq.~~\eqref{eq:cptpptm}. We refer the reader to Ref.~\cite{Helsen2019} for a detailed discussion of the spectral properties of the PTM of CPTP maps. 

Finally, we consider PTMs associated with commutators that act as $\mathcal{V}(A) = [V, A] = V A -A V$. The PTM takes the form

\begin{equation}
    \hat{\mathcal{V}} = \begin{pmatrix}
    0 & \bm{0}^T \\
    \bm{0} & \hat{\mathcal{V}}_{\Pi}
    \end{pmatrix}.
\end{equation}

\subsubsection{Single-qubit amplitude damping noise}
\label{app:subsubsecad}

Single-qubit amplitude damping noise $\mathcal{A}$ is a Lindbladian noise channel characterized by two Kraus operators $A_0$ and $A_1$ given by

\begin{equation}
    A_0 = \begin{pmatrix}
        1 & 0 \\
        0 & \sqrt{1- \gamma_{\downarrow}}
    \end{pmatrix}, \quad     A_1 = \begin{pmatrix}
        0 & \sqrt{\gamma_{\downarrow}} \\
        0 & 0
    \end{pmatrix},
\end{equation}
with the damping parameter $\gamma_{\downarrow} \in [0, 1]$.
Thus, the action of single-qubit amplitude damping noise on a density matrix is given by 

\begin{equation}
    \mathcal{A}(\rho) = A_0 \rho A_0^{\dagger} + A_1 \rho A_1^{\dagger}.
\end{equation}
Accordingly, the PTM reads

\begin{equation}
\label{eq:ptmad}
    \hat{\mathcal{A}} = \begin{pmatrix}
    1 & 0 & 0 & 0 \\
    0 & \sqrt{1-\gamma_{\downarrow}} & 0 & 0 \\
    0 & 0 & \sqrt{1-\gamma_{\downarrow} } & 0 \\
    \gamma_{\downarrow} & 0 & 0 & 1- \gamma_{\downarrow}
    \end{pmatrix}.
\end{equation}
In the main text, we considered single-qubit amplitude damping noise that acts independently on each qubit, whose PTM is simply $\hat{\mathcal{A}}^{\otimes n}$.

\section{Decay of the variance of the gradient in the toy model}
\label{app:var_grad_decay}

In this Appendix, we derive the analytical formulas reported in Subsec.~\ref{subsec:nibp2design} for our toy model of unitary $2$-design alternated with noise maps. We obtain the results using the Liouville superoperator formalism described in Appendix~\ref{app:preliminary}. 

Without loss of generality we can assume that the unitary $U(\bm{\theta}_{\ell})$ can be written as

\begin{equation}
U(\bm{\theta}_{\ell}) = \prod_{k=1}^K e^{-i \theta_{\ell k} V_{k} } W_k,
\end{equation}
where $V_k$ are Hermitian, traceless operators, and $W_k$ are fixed unitaries. In Liouville representation this becomes

\begin{equation}
\hat{\mathcal{U}}(\bm{\theta}_{\ell}) = \prod_{k=1}^K e^{-i \theta_{\ell k} \hat{\mathcal{V}}_{k} } \hat{\mathcal{W}}_k,
\end{equation}
where $\hat{\mathcal{V}}_k$ is the Liouville superoperator that represents the commutator that acts on a generic operator $A$ as $\mathcal{V}_k(A)= [V_k, A] = V_k A - A V_k$. 

We are interested in the average and the variance of the partial derivative of the cost $C(\bm{\theta})$, given in Eq.~\eqref{eq:costf}, with respect to the parameters $\theta_{\ell k}$. This can be written as 

\begin{equation}
\frac{\partial C}{\partial \theta_{\ell k}} = -i \bbra{O} \hat{\mathcal{N}} \hat{\mathcal{U}}(\bm{\theta}_L)  \dots \hat{\mathcal{N}}  \hat{\mathcal{U}}_{\ell+}(\bm{\theta}_{\ell +}) \hat{\mathcal{V}}_{k}  \hat{\mathcal{U}}_{\ell-}(\bm{\theta}_{\ell -}) \hat{\mathcal{N}} \dots \hat{\mathcal{N}} \hat{\mathcal{U}}(\bm{\theta}_1) \kket{\rho_{\mathrm{in}}},
\end{equation}
where we defined the unitaries

\begin{subequations}
\begin{align}
    \hat{\mathcal{U}}_{\ell+}(\bm{\theta}_{\ell +}) & = \prod_{j=k+1}^K e^{-i \theta_{\ell j} \hat{\mathcal{V}}_{ j}} \hat{\mathcal{W}}_{ j} ,  \\
    \hat{\mathcal{U}}_{\ell-}(\bm{\theta}_{\ell -}) & = \prod_{j=1}^k e^{-i \theta_{\ell j} \hat{\mathcal{V}}_{ j}} \hat{\mathcal{W}}_{ j}
\end{align}
\end{subequations}
with $\bm{\theta}_{\ell +} = \begin{pmatrix}
\theta_{\ell K} & \dots & \theta_{\ell (k+1)}
\end{pmatrix}$ and $\bm{\theta}_{\ell -} = \begin{pmatrix}
\theta_{\ell k} & \dots & \theta_{\ell 1}
\end{pmatrix}.$
Similar to Ref.~\cite{McClean2018}, the unitary $1$-design property suffices to obtain that $\forall k, \ell$, and for every possible cost function, we get

\begin{equation}
\biggl \langle \frac{\partial C}{\partial \theta_{\ell k}} \biggr \rangle = 0,
\end{equation}
where the average is taken over the uniform distribution in parameter space. 
Accordingly, we can write the variance as

\begin{multline}
\mathrm{Var}\biggl( \frac{\partial C}{\partial \theta_{\ell k}} \biggr) = \biggl \langle \biggl( \frac{\partial C}{\partial \theta_{\ell k}} \biggr)^2 \biggr \rangle  \\
= \int_{\mathcal{D}^L} d \bm{\theta}_L \dots d \bm{\theta}_{\ell +} d \bm{\theta}_{\ell -} \dots d \bm{\theta}_1 p(\bm{\theta}_{L}) \dots p_+(\bm{\theta}_{\ell +})p_-(\bm{\theta}_{\ell -}) \dots p(\bm{\theta}_{1})\times  \\
 \bbra{O} \hat{\mathcal{N}} \hat{\mathcal{U}}(\bm{\theta}_L) \dots \hat{\mathcal{N}} \hat{\mathcal{U}}_{\ell+}(\bm{\theta}_{\ell +}) \hat{\mathcal{V}}_{k}  \hat{\mathcal{U}}_{\ell-}(\bm{\theta}_{\ell -}) \hat{\mathcal{N}} \dots \hat{\mathcal{N}} \hat{\mathcal{U}}(\bm{\theta}_1) \kket{\rho_{\mathrm{in}}} \times  \\
   \bbra{\rho_{\mathrm{in}}} \hat{\mathcal{U}}^{\dagger}(\bm{\theta}_1) \hat{\mathcal{N}}^{\dagger} \dots \hat{\mathcal{N}}^{\dagger} \hat{\mathcal{U}}_{\ell-}^{\dagger}(\bm{\theta}_{\ell -}) \hat{\mathcal{V}}_{k} \hat{\mathcal{U}}_{\ell+}^{\dagger}(\bm{\theta}_{\ell +})\hat{\mathcal{N}}^{\dagger} \dots \hat{\mathcal{U}}^{\dagger}(\bm{\theta}_L) \hat{\mathcal{N}}^{\dagger} \kket{O},
\end{multline}
which we can rewrite as

\begin{multline}
\mathrm{Var}\biggl( \frac{\partial C}{\partial \theta_{ \ell k}} \biggr) = \int_{\mathcal{D}^L} d \bm{\theta}_L \dots d \bm{\theta}_{\ell +} d \bm{\theta}_{\ell -} \dots d \bm{\theta}_1 p(\bm{\theta}_{L}) \dots p_+(\bm{\theta}_{\ell +})p_-(\bm{\theta}_{\ell -}) \dots p(\bm{\theta}_{1})\times  \\
 \bbra{O} \hat{\mathcal{N}} \hat{\mathcal{U}}(\bm{\theta}_L) \dots \hat{\mathcal{N}} \hat{\mathcal{U}}_{\ell+} (\bm{\theta}_{\ell +}) \hat{\mathcal{U}}_{\ell-} (\bm{\theta}_{\ell -}) \hat{\mathcal{U}}_{\ell-}^{\dagger} (\bm{\theta}_{\ell -}) \hat{\mathcal{V}}_{k}  \hat{\mathcal{U}}_{\ell-}(\bm{\theta}_{\ell -}) \hat{\mathcal{N}} \dots \hat{\mathcal{N}} \hat{\mathcal{U}}(\bm{\theta}_1) \kket{\rho_{\mathrm{in}}} \times  \\
   \bbra{\rho_{\mathrm{in}}} \hat{\mathcal{U}}^{\dagger}(\bm{\theta}_1) \hat{\mathcal{N}}^{\dagger} \dots \hat{\mathcal{N}}^{\dagger} \hat{\mathcal{U}}_{\ell-}^{\dagger}(\bm{\theta}_{\ell -}) \hat{\mathcal{V}}_{k} \hat{\mathcal{U}}_{\ell-} (\bm{\theta}_{\ell -}) \hat{\mathcal{U}}_{\ell-}^{\dagger} (\bm{\theta}_{\ell -}) \hat{\mathcal{U}}_{\ell+}^{\dagger}(\bm{\theta}_{\ell +})\hat{\mathcal{N}}^{\dagger} \dots \hat{\mathcal{U}}^{\dagger}(\bm{\theta}_L) \hat{\mathcal{N}}^{\dagger} \kket{O} = \\
   \int_{\mathcal{D}^L} d \bm{\theta}_L \dots d \bm{\theta}_{\ell} d \bm{\theta}_{\ell -} \dots d \bm{\theta}_1 p(\bm{\theta}_{L}) \dots p(\bm{\theta}_{\ell})p_-(\bm{\theta}_{\ell -}) \dots p(\bm{\theta}_{1})\times  \\
 \bbra{O} \hat{\mathcal{N}} \hat{\mathcal{U}}(\bm{\theta}_L) \dots \hat{\mathcal{N}} \hat{\mathcal{U}} (\bm{\theta}_{\ell})  \hat{\mathcal{U}}_{\ell-}^{\dagger} (\bm{\theta}_{\ell -}) \hat{\mathcal{V}}_{k}  \hat{\mathcal{U}}_{\ell-}(\bm{\theta}_{\ell -}) \hat{\mathcal{N}} \dots \hat{\mathcal{N}} \hat{\mathcal{U}}(\bm{\theta}_1) \kket{\rho_{\mathrm{in}}} \times  \\
   \bbra{\rho_{\mathrm{in}}} \hat{\mathcal{U}}^{\dagger}(\bm{\theta}_1) \hat{\mathcal{N}}^{\dagger} \dots \hat{\mathcal{N}}^{\dagger} \hat{\mathcal{U}}_{\ell-}^{\dagger}(\bm{\theta}_{\ell -}) \hat{\mathcal{V}}_{k} \hat{\mathcal{U}}_{\ell-} (\bm{\theta}_{\ell -}) \hat{\mathcal{U}}^{\dagger} (\bm{\theta}_{\ell}) \hat{\mathcal{N}}^{\dagger} \dots \hat{\mathcal{U}}^{\dagger}(\bm{\theta}_L) \hat{\mathcal{N}}^{\dagger} \kket{O}.
\end{multline}
Since we assume that each layer forms a unitary $2$-design, we can substitute all the integrals, apart from the one over $\bm{\theta}_{\ell -}$ with Haar integrals to get

\begin{multline}
    \mathrm{Var}\biggl( \frac{\partial C}{\partial \theta_{ \ell k}} \biggr) 
    = \int d\mu(U_L) \dots d \mu (U_{\ell}) d \mu (U_{\ell - 1}) \dots d \mu(U_1) \int_{\mathcal{D_-}} d \bm{\theta}_{\ell -} p_-(\bm{\theta}_{\ell -}) \times  \\
 \bbra{O} \hat{\mathcal{N}} \hat{\mathcal{U}}_L \dots \hat{\mathcal{N}} \hat{\mathcal{U}}_{\ell} \hat{\mathcal{U}}_{\ell-}^{\dagger} (\bm{\theta}_{\ell -}) \hat{\mathcal{V}}_{k}  \hat{\mathcal{U}}_{\ell-}(\bm{\theta}_{\ell -}) \hat{\mathcal{N}} \dots \hat{\mathcal{N}} \hat{\mathcal{U}}_1 \kket{\rho_{\mathrm{in}}} \times  \\
   \bbra{\rho_{\mathrm{in}}} \hat{\mathcal{U}}^{\dagger}_1 \hat{\mathcal{N}}^{\dagger} \dots \hat{\mathcal{N}}^{\dagger} \hat{\mathcal{U}}_{\ell-}^{\dagger}(\bm{\theta}_{\ell -}) \hat{\mathcal{V}}_{k} \hat{\mathcal{U}}_{\ell-} (\bm{\theta}_{\ell -}) \hat{\mathcal{U}}_{\ell}^{\dagger} \hat{\mathcal{N}}^{\dagger} \dots \hat{\mathcal{U}}^{\dagger}_L \hat{\mathcal{N}}^{\dagger} \kket{O}
\end{multline}
Now we can iteratively apply Eq.~\eqref{eq:haar3} with different Liouville superoperators $\hat{\mathcal{M}}_j$, starting from

\begin{equation}
    \hat{\mathcal{M}}_1 = \kket{\rho_{\mathrm{in}}} \bbra{\rho_{\mathrm{in}}}.
\end{equation}
We obtain

\begin{equation}
\int d \mu(U_1) \hat{\mathcal{U}}_1 \mathcal{\hat{M}}_1 \hat{\mathcal{U}}_1^{\dagger} = \alpha_1 \kket{\pwn{0}} \bbra{\pwn{0}} + \beta_1 \hat{\Pi}, 
\end{equation}
where $\hat{\Pi}$ is the projector in Eq.~\eqref{eq:proj} and where we defined the coefficients 

\begin{subequations}
\label{eq:coeffin}
\begin{equation}
\alpha_1 = \frac{1}{2^n}, 
\end{equation} 
\begin{equation}
\beta_1 = \frac{\mathrm{Tr}(\rho_{\mathrm{in}}^2) - 1/2^n}{4^n -1}. 
\end{equation}
\end{subequations}
In the second iteration, we identify the superoperator

\begin{equation}
\hat{\mathcal{M}}_2 =  \alpha_1 \hat{\mathcal{N}} \kket{\pwn{0}} \bbra{\pwn{0}} \hat{\mathcal{N}}^{\dagger} + \beta_1 \hat{\mathcal{N}} \hat{\Pi} \hat{\mathcal{N}}^{\dagger} . 
\end{equation}
From the structure of the PTMs of $\hat{\mathcal{N}}$ and $\hat{\mathcal{N}}^{\dagger}$ given in Eq.~\eqref{eq:cptpptm} and ~\eqref{eq:cptpptmdag}, respectively, it follows importantly that

\begin{equation}
    \hat{\mathcal{N}} \hat{\Pi} \hat{\mathcal{N}}^{\dagger} = \hat{\Pi} \hat{\mathcal{N}} \hat{\mathcal{N}}^{\dagger} \hat{\Pi}  = \begin{pmatrix}
0 & \bm{0}^T \\
\bm{0} & \hat{\mathcal{N}}_{\Pi} \hat{\mathcal{N}}_{\Pi}^{\dagger}
\end{pmatrix}. 
\end{equation}
Before continuing we define the ``noise'' coefficients similar to Ref.~\cite{eisert} as

\begin{subequations}
\label{eq:noisecoeff}
\begin{equation}
\label{eq:nun}
    \nu_{\mathcal{N}} = \frac{\mathrm{Tr}(\hat{\mathcal{N}} \hat{\mathcal{N}}^{\dagger})}{4^n} = \frac{\mathrm{Tr}[I \otimes \mathcal{N} (\ket{\Gamma}\bra{\Gamma})^2]}{4^n},
\end{equation}

\begin{equation}
    \label{eq:eta}
\eta_{\mathcal{N}} \equiv \frac{\mathrm{Tr}(\hat{\mathcal{N}}^{\dagger} \hat{\mathcal{N}} \kket{\pwn{0}}\bbra{\pwn{0}})}{2^n} = \mathrm{Tr}\biggl[\mathcal{N}\biggl(\frac{I}{2^n} \biggr)^2 \biggr],
\end{equation}

\begin{equation}
\label{eq:rn}
r_{\mathcal{N}} \equiv \frac{\mathrm{Tr}(\hat{\mathcal{N}} \hat{\Pi} \hat{\mathcal{N}}^{\dagger})}{4^n -1} = \frac{\mathrm{Tr}(\hat{\mathcal{N}}_{\Pi}  \hat{\mathcal{N}}_{\Pi}^{\dagger})}{4^n -1} = \frac{4^n \nu_{\mathcal{N}} - 2^n \eta_{\mathcal{N}}}{4^n-1},
\end{equation} 
\end{subequations}
where $\ket{\Gamma} = \sum_{i=0}^{2^n-1}\ket{i, i}$ is the unnormalized maximally entangled state between the system and a copy of itself.  We see that $\eta_{\mathcal{N}}$ is simply the purity of the state obtained by applying the noise map $\mathcal{N}$ to the completely mixed state, and thus it takes values in $[1/2^n, 1]$. Additionally, Ref.~\cite{eisert} (see Proposition $2$ in Supplementary Material VII) proved that $0 < r_{\mathcal{N}} \le 1$.  

Thus, using again Eq.~\eqref{eq:haar3} we compute the integrals

\begin{subequations}
\label{eq:noise_int}
\begin{equation}
\int d \mu(U_2) \hat{\mathcal{U}}_2 \hat{\mathcal{N}} \kket{\pwn{0}} \bbra{\pwn{0}} \hat{\mathcal{N}}^{\dagger} \hat{\mathcal{U}}_2^{\dagger} = \frac{2^n \eta_{\mathcal{N}} - 1}{4^n - 1} \hat{\Pi} +  \kket{\pwn{0}} \bbra{\pwn{0}},
\end{equation}

\begin{equation}
\label{eq:noise_int_pi}
\int  d \mu (U_2) \hat{\mathcal{U}}_2 \hat{\mathcal{N}} \hat{\Pi} \hat{\mathcal{N}}^{\dagger} \hat{\mathcal{U}}_2^{\dagger} =r_{\mathcal{N}} \hat{\Pi}.
\end{equation}
\end{subequations}
In obtaining the previous equations we used that

\begin{equation}
    \mathrm{Tr}(\hat{\mathcal{N}} \hat{\mathcal{N}}^{\dagger} \kket{\pwn{0}}\bbra{\pwn{0}}) = 1,
\end{equation}
which follows again from the structure of the PTM of $\hat{\mathcal{N}}$ and $\hat{\mathcal{N}}^{\dagger}$.

We obtain

\begin{equation}
\int  d \mu (U_2) \hat{\mathcal{U}}_2 \hat{\mathcal{M}_2} \hat{\mathcal{U}}_2^{\dagger} = \alpha_2 \kket{\pwn{0}} \bbra{\pwn{0}} + \beta_2 \hat{\Pi},
\end{equation}
with coefficients

\begin{subequations}
\begin{align}
\alpha_2 &=  \frac{1}{2^n}, \\
\beta_2 &= \frac{\eta_{\mathcal{N}} -1/2^n}{4^n-1} + r_{\mathcal{N}} \beta_1. 
\end{align}
\end{subequations}
We can keep on computing integrals in this way till layer $\ell - 1$. In particular, at iteration $j$ we consider the Liouville superoperator

\begin{equation}
\label{eq:mit}
    \hat{\mathcal{M}}_j = \alpha_{j-1} \hat{\mathcal{N}} \kket{\pwn{0}} \bbra{\pwn{0}} \hat{\mathcal{N}}^{\dagger} + 
    \beta_{j-1} \hat{\mathcal{N}} \hat{\Pi} \hat{\mathcal{N}}^{\dagger}.
\end{equation}
and obtain new coefficients

\begin{subequations}
\label{eq:coeffit}
\begin{align}
\alpha_j &= \frac{1}{2^n}, \\
\beta_j &= \frac{\eta_{\mathcal{N}} -1/2^n}{4^n-1} + r_{\mathcal{N}} \beta_{j-1}. 
\end{align}
\end{subequations}
for $j=2, \dots, \ell -1 $ and starting the iteration from the coefficients $\alpha_{1}, \beta_1$ in Eqs.~\eqref{eq:coeffin}.

We arrive at

\begin{multline}
\mathrm{Var}\biggl( \frac{\partial C}{\partial \theta_{\ell k}} \biggr) 
=  \int d \mu (U_L) \dots d \mu(U_{\ell}) \int_{\mathcal{D_-}} d \bm{\theta}_{\ell -} p_-(\bm{\theta}_{\ell -})  \times \\ \bbra{O} \hat{\mathcal{N}} \hat{\mathcal{U}}_L \dots \hat{\mathcal{U}}_{\ell}  \hat{\mathcal{V}}_{\ell k}^{-}(\bm{\theta}_{\ell -}) \hat{\mathcal{M}}_{\ell-1} 
\hat{\mathcal{V}}_{\ell k}^{-}(\bm{\theta}_{\ell -}) \hat{\mathcal{U}}_{\ell}^{\dagger}  \dots \hat{\mathcal{U}}_L^{\dagger} \hat{\mathcal{N}}^{\dagger} \kket{O},
\end{multline}
where we defined the Hermitian superoperator 

\begin{equation}
    \hat{\mathcal{V}}_{\ell k}^{-}(\bm{\theta}_{\ell -}) = \hat{\mathcal{U}}_{\ell-}^{\dagger}(\bm{\theta}_{\ell -}) \hat{\mathcal{V}}_{ k}  \hat{\mathcal{U}}_{\ell-}(\bm{\theta}_{\ell -}).
\end{equation}
Thus, at layer $\ell$ we can apply the Haar integral to the superoperator

\begin{multline}
    \hat{\mathcal{M}}_{\ell}(\bm{\theta}_{\ell-}) = \hat{\mathcal{V}}_{\ell k}^{-}(\bm{\theta}_{\ell -}) \hat{\mathcal{M}}_{\ell-1} 
\hat{\mathcal{V}}_{\ell k}^{-}(\bm{\theta}_{\ell -}) = \\
 \alpha_{\ell-1} \hat{\mathcal{V}}_{\ell k}^{-}(\bm{\theta}_{\ell -}) \hat{\mathcal{N}} \kket{\pwn{0}} \bbra{\pwn{0}} \hat{\mathcal{N}}^{\dagger} \hat{\mathcal{V}}_{\ell k}^{-}(\bm{\theta}_{\ell -}) + \beta_{\ell -1} \hat{\mathcal{V}}_{\ell k}^{-}(\bm{\theta}_{\ell -}) \hat{\mathcal{N}} \hat{\Pi} \hat{\mathcal{N}}^{\dagger} \hat{\mathcal{V}}_{\ell k}^{-}(\bm{\theta}_{\ell -}).
\end{multline}
We get

\begin{subequations}
\begin{multline}
    \int d \mu(U_{\ell}) \hat{\mathcal{U}}_{\ell} \hat{\mathcal{V}}_{\ell k}^{-}(\bm{\theta}_{\ell -}) \hat{\mathcal{N}} \kket{\pwn{0}} \bbra{\pwn{0}} \hat{\mathcal{N}}^{\dagger} \hat{\mathcal{V}}_{\ell k}^{-}(\bm{\theta}_{\ell -}) \hat{\mathcal{U}_{\ell}}^{\dagger} = \\
    \frac{\mathrm{Tr}[\hat{\mathcal{V}}_{\ell k}^{-}(\bm{\theta}_{\ell -}) \hat{\mathcal{N}} \kket{\pwn{0}}\bbra{\pwn{0}} \hat{\mathcal{N}}^{\dagger} \hat{\mathcal{V}}_{\ell k}^{-}(\bm{\theta}_{\ell -}) \hat{\Pi} ] }{4^n-1} \hat{\Pi},
\end{multline}

\begin{equation}
    \int d \mu(U_{\ell}) \hat{\mathcal{U}}_{\ell} \hat{\mathcal{V}}_{\ell k}^{-}(\bm{\theta}_{\ell -}) \hat{\mathcal{N}} \hat{\Pi} \hat{\mathcal{N}}^{\dagger} \hat{\mathcal{V}}_{\ell k}^{-}(\bm{\theta}_{\ell -}) \hat{\mathcal{U}_{\ell}}^{\dagger} = \frac{\mathrm{Tr}[\hat{\mathcal{V}}_{\ell k}^{-}(\bm{\theta}_{\ell -}) \hat{\mathcal{N}} \hat{\mathcal{N}}^{\dagger} \hat{\mathcal{V}}_{\ell k}^{-}(\bm{\theta}_{\ell -}) \hat{\Pi} ] }{4^n-1} \hat{\Pi}.
\end{equation}
\end{subequations}
We define the average coefficients over the $\bm{\theta}_{\ell -}$ parameters

\begin{subequations}
\begin{equation}
    \langle \eta_{\ell k}^{(-)} \rangle = \int_{\mathcal{D-}} d \bm{\theta_{\ell-}} p(\bm{\theta}_{\ell-}) \mathrm{Tr}[\hat{\mathcal{V}}_{\ell k}^{-}(\bm{\theta}_{\ell -}) \hat{\mathcal{N}} \kket{\pwn{0}}\bbra{\pwn{0}} \hat{\mathcal{N}}^{\dagger} \hat{\mathcal{V}}_{\ell k}^{-}(\bm{\theta}_{\ell -}) \hat{\Pi} ],
\end{equation}

\begin{equation}
    \langle \nu_{\ell k}^{(-)} \rangle = \int_{\mathcal{D-}} d \bm{\theta_{\ell-}} p(\bm{\theta}_{\ell-}) \mathrm{Tr}[\hat{\mathcal{V}}_{\ell k}^{-}(\bm{\theta}_{\ell -}) \hat{\mathcal{N}} \hat{\mathcal{N}}^{\dagger} \hat{\mathcal{V}}_{\ell k}^{-}(\bm{\theta}_{\ell -}) \hat{\Pi} ],
\end{equation}
\end{subequations}
which allow us to write compactly the variance as

\begin{multline}
    \mathrm{Var}\biggl( \frac{\partial C}{\partial \theta_{\ell k}} \biggr) 
= \biggl(\frac{\langle \eta_{\ell k}^{(-)} \rangle}{2^n(4^n-1)} + \frac{\langle \nu_{\ell k}^{(-)} \rangle}{4^n-1}\beta_{\ell - 1} \biggr) \times \\
\int d \mu(U_L) \dots d \mu(U_{\ell+1}) \bbra{O} \hat{\mathcal{N}}  \hat{\mathcal{U}}_L \dots \hat{\mathcal{U}}_{\ell +1} \hat{\mathcal{N}}  \hat{\Pi}\hat{\mathcal{N}}^{\dagger} \hat{\mathcal{U}}_{\ell+1}^{\dagger} \dots \hat{\mathcal{U}}_{L}^{\dagger} \hat{\mathcal{N}}^{\dagger} \kket{O}.
\end{multline}
Now we can simply continue with the iterative Haar integration of the projector $\hat{\Pi}$ using Eq.~\eqref{eq:noise_int_pi} to get the final formula for the variance

\begin{equation}
\mathrm{Var}\biggl( \frac{\partial C}{\partial \theta_{\ell k}} \biggr) = \biggl(\frac{\langle \eta_{\ell k}^{(-)} \rangle}{2^n} + \langle \nu_{\ell k}^{(-)} \rangle\beta_{\ell - 1} \biggr) \frac{r_{\mathcal{N}}^{L - \ell}}{4^n-1} \mathrm{Tr}(\mathcal{N}^{\dagger}(O)^2),
\end{equation}
which defining the coefficient

\begin{equation}
\label{eq:gkl}
    G_{{\ell k}} =  \biggl(\frac{\langle \eta_{\ell k}^{(-)} \rangle}{2^n} + \langle \nu_{\ell k}^{(-)} \rangle\beta_{\ell - 1} \biggr)
\end{equation}
gives the result in Eq.~\eqref{eq:var_decay}. 

The result we obtained is valid for any CPTP map. We now further discuss specific cases. If the noise is unital, i.e., $\mathcal{N}(I) = I$, then $\eta_{\mathcal{N}} = 1/2^n$ and $\langle \eta_{\ell k}^{(-)} \rangle = 0$. In this case, the general formulas  simplifies to Eq.~\eqref{eq:var_decay_unital} in the main text. 

For weak, Lindbladian noise maps, we take a first order approximation in the noise strength as in Eq. ~\eqref{eq:weakmarkoviannoise}. Using this approximation, and keeping only first order terms in the noise strength, we can approximate the noise coefficients as 

\begin{subequations}
\label{eq:approxcoeffmarkovian}
\begin{equation}
    \eta_{\mathcal{N}} \approx \frac{1}{2^n},
\end{equation}
\begin{equation}
    \nu_{\mathcal{N}} \approx 1 + 2 \sum_{k} \frac{\mathrm{Tr}(\hat{\mathcal{D}}[L_k])}{4^n}.
\end{equation}
\end{subequations}
Notice that the approximation for $\eta_{\mathcal{N}}$ states that Linbladian noise is unital up to first order in the noise strength. Using Eq.~\eqref{eq:trdjump}, we can approximate the noise coefficient $\nu_{\mathcal{N}}$ as 

\begin{equation}
\label{eq:approxnumarkovian}
    \nu_{\mathcal{N}} \approx 1 - \frac{1}{2^{n-1}} \sum_k \mathrm{Tr} \bigl[L_{k}^{\dagger} L_k \bigr] \le 1,
\end{equation}
and substituting into the definition of the coefficient $r_{\mathcal{N}}$ in Eq.~\eqref{eq:rn} we get Eq.~\eqref{eq:rnmarkovian} in the main text. 

\section{Analysis of the overlap between quantum states in the toy model} 
\label{sec:app_c}

\subsection{Average overlap}
\label{subsec:avgoverlap}
In this section, we consider the following problem. We have two quantum states $\rho_{ i}$ and $\rho_{j}$. The states are sent to the same quantum channel that is given by a random unitary $U$ sampled from a unitary $2$-design and a CPTP map $\mathcal{N}$. The output states are $\rho_{\mathrm{out}, i( j)} = \mathcal{N} \circ \mathcal{U}(\rho_{ i (j)})$. We ask ourselves what is the average overlap $\mathrm{Tr}(\rho_{\mathrm{out}, i} \ \rho_{\mathrm{out}, j})$  over all possible $U$, that is

\begin{equation}
\label{eq:avgoverlap}
\mathbb{E}_{U} \mathrm{Tr}(\rho_{\mathrm{out}, i} \rho_{\mathrm{out}, j}) = \int d \mu(U) \bbra{\rho_{i}} \hat{\mathcal{U}}^{\dagger} \hat{\mathcal{N}}^{\dagger} \hat{\mathcal{N}} \hat{\mathcal{U}} \kket{\rho_{j}} = 
\bbra{\rho_{i}} \biggl( \int d \mu(U) \hat{\mathcal{U}}^{\dagger} \hat{\mathcal{N}}^{\dagger} \hat{\mathcal{N}} \hat{\mathcal{U}}  \biggr)\kket{\rho_{j}}. 
\end{equation}
This is the same problem treated in Ref.~\cite{eisert}, and in what follows we just derive their result using the Liouville superoperator formalism based on the application of Eq.~\eqref{eq:haar3}. Notice that if we take $\rho_{i} = \rho_{j}$ we get the average purity. Looking at Eq.~\eqref{eq:avgoverlap} we see that all we need is to use Eq.~\eqref{eq:haar3} with $\hat{\mathcal{M}} = \hat{\mathcal{N}}^{\dagger} \hat{\mathcal{N}}$. After some algebra we get

\begin{equation}
\label{eq:avgoverlap2}
\mathbb{E}_{U} \mathrm{Tr}(\rho_{\mathrm{out}, i} \rho_{\mathrm{out}, j}) = r_{\mathcal{N}} \mathrm{Tr}[\rho_i \rho_j ] + \frac{2^n}{4^n-1}(2^n \eta_{\mathcal{N}} - \nu_{\mathcal{N}}),
\end{equation}
where the coefficients $\eta_{\mathcal{N}}$, $\nu_{\mathcal{N}}$ and $r_{\mathcal{N}}$ are the noise coefficients given in Eqs.~\eqref{eq:noisecoeff}.  

From a first inspection of Eq.~\eqref{eq:avgoverlap2} it is clear that we do not expect the average overlap to be the same as the overlap after the application of the average Haar random channel associated with $\hat{\mathcal{N}}$ given in Eq.~\eqref{eq:twirl1}. In fact, $\hat{\mathcal{N}}_{\mathrm{Haar}}$ is completely determined by $p_{\mathrm{eff}}$ given in Eq.~\eqref{eq:ptwirl}, which does not seem to be immediately related to  $\nu_{\mathcal{N}}$ nor $\eta_{\mathcal{N}}$. The overlap after the application of the Haar averaged channel is 

\begin{equation}
\label{eq:avgtwirloverlap}
\mathrm{Tr}[\mathcal{N}_{\mathrm{Haar}}(\rho_i) \mathcal{N}_{\mathrm{Haar}}(\rho_j)] = (1 - p_{\mathrm{eff}})^2 \mathrm{Tr}[\rho_i \rho_j] +  \frac{2 p_{\mathrm{eff}} (1-p_{\mathrm{eff}})}{2^n} + \frac{p_{\mathrm{eff}}^2}{2^n} 
\end{equation}
If we now consider $L$ layers and apply Eq.~\eqref{eq:avgoverlap2} iteratively we get the same result as Eq.\:(173) in the Supplementary Material of Ref.~\cite{eisert} that reads

\begin{equation}
\label{eq:avgoverlap2_all}
    \mathbb{E}_{U_L \dots U_1} \mathrm{Tr}(\rho_{\mathrm{out}, i} \rho_{\mathrm{out}, j}) = r_{\mathcal{N}}^L \mathrm{Tr}(\rho_i \rho_j) + \frac{1 - r_{\mathcal{N}}^L}{1 - r_{\mathcal{N}}} \frac{2^n (2^n \eta_{\mathcal{N}} - \nu_{\mathcal{N}})}{4^n-1}. 
\end{equation}
Thus, when $r_{\mathcal{N}} < 1$ the average overlap between any two initial states subject to the repeated application of random unitaries and noise channel goes asymptotically to

\begin{multline}
\label{eq:asymptoticpurity}
    \mathbb{E}_{L \rightarrow \infty}\mathrm{Tr}(\rho_{\mathrm{out}, i} \rho_{\mathrm{out}, j}) \equiv \lim_{L \rightarrow  \infty }  \mathbb{E}_{U_L \dots U_1} \mathrm{Tr}(\rho_{\mathrm{out}, i} \rho_{\mathrm{out}, j}) = \\
    \frac{1}{1 - r_{\mathcal{N}}}
    \frac{2^n (2^n \eta_{\mathcal{N}} - \nu_{\mathcal{N}})}{4^n-1} = 
    \frac{2^n (2^n \eta_{\mathcal{N}} - \nu_{\mathcal{N}})}{4^n-1 + 2^n (\eta_{\mathcal{N}} -2^n \nu_{\mathcal{N}})}.
\end{multline} 

It is interesting to study the behavior of the average purity for specific kinds of noise, and in particular local noise models. Let us consider for instance the local amplitude damping on each qubit with parameter $\gamma_{\downarrow}$ described in Appendix~\ref{app:subsubsecad}. This gives the noise coefficients 

\begin{subequations}
\begin{equation}
    \eta_{\mathcal{A}} = \frac{(1 + \gamma_{\downarrow}^2)^n}{2^n},
\end{equation}
\begin{equation}
    \nu_{\mathcal{A}} = \frac{[2 + (\gamma_{\downarrow} - 2)\gamma_{\downarrow}]^n}{2^n}.
\end{equation}
\begin{equation}
    r_{\mathcal{A}} = \frac{2^n [2 + (\gamma_{\downarrow} - 2)\gamma_{\downarrow}]^n  -(1 + \gamma_{\downarrow}^2)^n}{4^n-1}.
\end{equation}
\end{subequations}
Plugging these formulas into Eq.~\eqref{eq:asymptoticpurity} we get that for local amplitude damping the average asymptotic overlap is given by

\begin{equation}
\label{eq:asymptoticpurityad}
    \mathbb{E}_{L \rightarrow \infty}^{(\mathcal{A})}\mathrm{Tr}(\rho_{\mathrm{out}, i} \rho_{\mathrm{out}, j})   =  \frac{1}{1-\frac{2^n [2 + (\gamma_{\downarrow} - 2)\gamma_{\downarrow}]^n  -(1 + \gamma_{\downarrow}^2)^n}{4^n-1}}  \frac{2^n(1 + \gamma_{\downarrow}^2)^n - [2 + (\gamma_{\downarrow} - 2)\gamma_{\downarrow}]^n}{4^n-1}.
\end{equation}
Now for any $0<\gamma_{\downarrow}<1$ we get that in the large number of qubit limit

\begin{multline}
    \lim_{n \rightarrow + \infty} \mathbb{E}_{L \rightarrow \infty}^{(\mathcal{A})}\mathrm{Tr}(\rho_{\mathrm{out}, i} \rho_{\mathrm{out}, j}) = \\
    \lim_{n \rightarrow  \infty} \frac{1}{1-\frac{2^n [2 + (\gamma_{\downarrow} - 2)\gamma_{\downarrow}]^n  -(1 + \gamma_{\downarrow}^2)^n}{4^n-1}}  \frac{2^n(1 + \gamma_{\downarrow}^2)^n - [2 + (\gamma_{\downarrow} - 2)\gamma_{\downarrow}]^n}{4^n-1}  =  0.
\end{multline}
This shows that for local amplitude damping for large number of qubits and in the limit of high depth the average overlap between two initial states coincides with the purity of the maximally mixed state. This also suggests that in this limit any initial state will approach the completely mixed state. 

\subsubsection{Average overlap for weak Lindbladian noise }
\label{subapp:avgoverlapmarkovian}
We now specify our results to general dissipative Lindbladian noise described in Appendix.~\ref{subapp:markovian} in the weak noise limit. To first order in the noise strength the approximate coefficients $\eta_{\mathcal{N}}$ and $\nu_{\mathcal{N}}$ are given in Eqs.~\eqref{eq:approxcoeffmarkovian}, ~\eqref{eq:approxnumarkovian}. Additionally, the $p_{\mathrm{eff}}$ in the corresponding twirled channel given in Eq.~\eqref{eq:ptwirl}, within the same first order approximation reads

\begin{equation}
p_{\mathrm{eff}} \approx  -\sum_k \frac{\mathrm{Tr}(\hat{\mathcal{D}}[L_k])}{4^n-1} = \frac{2^n}{4^n-1} \sum_k \mathrm{Tr}[L_k^{\dagger} L_k].
\end{equation}
Plugging these approximations into Eq.~\eqref{eq:avgoverlap2} and Eq.~\eqref{eq:avgtwirloverlap} and keeping only up to first order terms we get that

\begin{multline}
\mathbb{E}_{U} \mathrm{Tr}(\rho_{\mathrm{out}, i} \rho_{\mathrm{out}, j}) \approx \mathrm{Tr}[\mathcal{N}_{\mathrm{Haar}}(\rho_i) \mathcal{N}_{\mathrm{Haar}}(\rho_j)] \approx \\ 
\mathrm{Tr}[\rho_i \rho_j] - \frac{2^{n+1}}{4^n-1} \biggl( \mathrm{Tr}[\rho_i \rho_j] - \frac{1}{2^n} \biggr) \sum_k \mathrm{Tr}(L_k^{\dagger} L_k). \label{eq:haareqalavg} 
\end{multline}
Thus, for weak Lindbladian noise maps the average overlap between two states after application of a unitary $2$-design and noise matches the overlap between these two states after application of the Haar averaged channel. Note that at the completely mixed state we get $\mathrm{Tr}[\rho_i \rho_j] - 1/2^n  =0$, which means that the result only makes sense if the purity of the states is much larger than the purity corresponding to the completely mixed state. Assuming $\mathrm{Tr}[\rho_i \rho_j]  \gg 1/2^n$ results in

\begin{multline}
\mathbb{E}_{U} \mathrm{Tr}(\rho_{\mathrm{out}, i} \rho_{\mathrm{out}, j}) \approx \mathrm{Tr}[\mathcal{N}_{\mathrm{Haar}}(\rho_i) \mathcal{N}_{\mathrm{Haar}}(\rho_j)] \approx \\ 
\biggl(1 - \frac{2^{n+1}}{4^n-1} \sum_k \mathrm{Tr}(L_k^{\dagger} L_k) \biggr)\mathrm{Tr}[\rho_i \rho_j] \approx (1 - 2 p_{\mathrm{eff}}) \mathrm{Tr}[\rho_i \rho_j].
\end{multline}
We further note that in the large number of qubits limit

\begin{equation}
    1-2 p_{\mathrm{eff}} \overset{n \gg 1}{\approx} 1 -\frac{1}{2^{n-1}} \sum_{k} \mathrm{Tr}(L_k L_k^{\dagger}), 
\end{equation}
which coincides within the same approximation with the coefficient $r_{\mathcal{N}}$ given in Eq.~\eqref{eq:rnmarkovian}. This shows that for weak Lindbladian noise maps, the decay of the variance of the gradient is controlled by approximately the same parameter that controls the decay of the average overlap, and thus also the purity. 

Finally, let us also obtain the asymptotic average overlap to first order for weak Lindbladian noise maps. We get that

\begin{equation}
    \mathbb{E}_{L \rightarrow \infty}\mathrm{Tr}(\rho_{\mathrm{out}, i} \rho_{\mathrm{out}, j}) \approx
    \frac{4^n -1}{2^{n+1}} \frac{1}{\sum_{k} \mathrm{Tr}[L_k^{\dagger} L_k]}
    \frac{2}{4^n-1} \sum_{k} \mathrm{Tr}[L_k^{\dagger} L_k]
    = \frac{1}{2^n}.
\end{equation}

\subsection{Single instance purity in the weak noise limit}
\label{app:pur_single}

We now want to go beyond the average description obtained in the last subsection and study the behavior of the single instances. In this subsection, we focus solely on the purity. In particular, we consider single runs of the toy model circuit with randomly chosen unitary $2$-designs at each layer. To study the evolution of the purity along the circuit, let us define the relative purity change per layer 

\begin{equation}
    \xi_{\ell} = \frac{\mathrm{Tr}(\rho_{\ell}^2)}{\mathrm{Tr}(\rho_{\ell-1}^2)}, \quad \ell = 1, \dots, L,
\end{equation}
with $\rho_{\ell'}$ the state at layer $\ell'=0, \dots, L$.
Note that we assume $\rho_{0}=\rho_{\mathrm{in}}$ so $\mathrm{Tr}(\rho_0^2)$ is the initial purity.  The purity at layer $\ell$ can be obtained by tracking the purity changes over all layers

\begin{equation}
\label{eq:pur_rhol}
    \mathrm{Tr}(\rho_{\ell}^2) = \mathrm{Tr}(\rho_{0}^2) \prod_{\ell'=1}^{\ell}   \xi_{\ell'} .
\end{equation}

 Thus, to describe the statistical behavior of $\mathrm{Tr}(\rho_{\ell}^2)$, we need to understand the random process undergone by the random variable $\xi_{\ell}$. As can be seen  from Eq.~\eqref{eq:haar_one_layer}, in the small noise limit, and as long as the purity is much larger than $1/2^n$, the average of $\xi_{\ell}$ is $\braket{\xi_{\ell}} \approx 1-2 p_{\mathrm{eff}} = \langle \xi \rangle$, independently of the layer $\ell$. Sampling from a random variable with constant average, for deep circuits with $\ell$ layers the arithmetic average $\overline{\xi}_a$ of $\xi_{\ell'}$ converges to the true average 
 
 \begin{equation}
     \overline{\xi}_a = \frac{1}{\ell} \sum_{\ell'=1}^{\ell} \xi_{\ell'} \overset{\ell \gg 1}{ \rightarrow}\langle \xi \rangle .
 \end{equation} 
 This can be formalized by means Hoeffding's inequality that describes the maximal distance between the sample average and the real average in a probabilistic way  \cite{hoeffding1963}

\begin{equation}
\label{eq:hoeffding}
   \mathrm{Prob} \biggl( \bigl \lvert \overline{\xi}_a - \langle \xi \rangle \bigr \rvert \geq \Delta \biggr) =  \mathrm{Prob} \biggl( \biggl \lvert \frac{1}{\ell} \sum_{\ell'=1}^{\ell}  \xi_{\ell'}   - \langle \xi \rangle \biggr \rvert \geq \Delta     \biggr) \leq 2 \exp \biggl[\frac{-2 \ell \Delta^2}{R^2} \biggr],
\end{equation}
where on the left-hand side we have the probability that the arithmetic average deviates from the real average by more than a maximal allowed deviation $\Delta$. The parameter $R$ describes the range of the random variables $\xi_{\ell'}$. The range $R$ can be bounded using the eigenvalues $\hat{\mathcal{N}}^{\dagger} \hat{\mathcal{N}}$ as we will now show. The purity at the layer $\ell$ can be written as

\begin{equation}
\label{eq:purityl}
    \mathrm{Tr}(\rho_{\ell}^2) = \bbra{\rho_{\ell-1}} \hat{\mathcal{U}} \hat{\mathcal{N}}^{\dagger} \hat{\mathcal{N}} \hat{\mathcal{U}} \kket{\rho_{\ell -1}}.
\end{equation}
Notice that $\kket{\rho_{\ell-1}}$ is not a normalized vector, but we can normalize it by dividing it by its norm $\lVert \kket{\rho_{\ell -1}} \rVert = \sqrt{\mathrm{Tr}(\rho_{\ell -1}^2)}$, and define the unit vector $\kket{\overline{\rho}_{\ell-1}} = \kket{\rho_{\ell-1}}/\sqrt{\mathrm{Tr}(\rho_{\ell-1}^2)}$. In this way we can rewrite Eq.~\eqref{eq:purityl} as

\begin{equation}
    \mathrm{Tr}(\rho_{\ell}^2) = \mathrm{Tr}(\rho_{\ell-1}^2) \bbra{\overline{\rho}_{\ell-1}} \hat{\mathcal{U}} \hat{\mathcal{N}}^{\dagger} \hat{\mathcal{N}} \hat{\mathcal{U}} \kket{\overline{\rho}_{\ell -1}}.
\end{equation}
Thus, we get that the maximum and minimum values of the random variables $\xi_{\ell}$ is 

\begin{align}
\xi_{\ell}^{\mathrm{max}} = &  \underbrace{\mathrm{max}}_{\sigma \in \mathrm{D}_n: \mathrm{Tr}(\sigma^2)=\mathrm{Tr}(\rho_{\ell-1}^2)} \bbra{\overline{\sigma}}  \hat{\mathcal{N}}^{\dagger}\hat{\mathcal{N}} \kket{\overline{\sigma}}, \\
\xi_{\ell}^{\mathrm{min}} = &  \underbrace{\mathrm{min}}_{\sigma \in \mathrm{D}_n: \mathrm{Tr}(\sigma^2)=\mathrm{Tr}(\rho_{\ell-1}^2)} \bbra{\overline{\sigma}}  \hat{\mathcal{N}}^{\dagger}\hat{\mathcal{N}} \kket{\overline{\sigma}}.
\end{align}
To remove the dependency on the purity at layer $\ell-1$ we can extend the maximization and minimization over all unit vectors to write

\begin{align}
\xi_{\ell}^{\mathrm{max}} \le &  \underbrace{\mathrm{max}}_{\overline{\sigma} :\bbrakket{\overline{\sigma}}{\overline{\sigma}}=1} \bbra{\overline{\sigma}}  \hat{\mathcal{N}}^{\dagger}\hat{\mathcal{N}} \kket{\overline{\sigma}} = \lambda_{\mathrm{max}}, \\
\xi_{\ell}^{\mathrm{min}} \ge &  \underbrace{\mathrm{min}}_{\overline{\sigma} :\bbrakket{\overline{\sigma}}{\overline{\sigma}}=1} \bbra{\overline{\sigma}}  \hat{\mathcal{N}}^{\dagger}\hat{\mathcal{N}} \kket{\overline{\sigma}} = \lambda_{\mathrm{min}},
\end{align}
where  $\lambda_{\mathrm{max}},\lambda_{\mathrm{min}} \ge 0$ are the maximum and minimum eigenvalue of the positive semidefinite superoperator $\hat{\mathcal{N}}^{\dagger} \hat{\mathcal{N}}$, respectively. Thus, we can take as range of the random variable $\xi_{\ell}$

\begin{equation}
\label{eq:range_r}
    R = \lambda_{\mathrm{max}} - \lambda_{\mathrm{min}}.
\end{equation}
Eq.~\eqref{eq:hoeffding} enables us to describe the convergence of the arithmetic average to the true average. However, as we see from Eq.~\eqref{eq:pur_rhol}, the purity at layer $\ell$ is not connected to the arithmetic average of the random variable $\xi_{\ell'}$, but rather to the geometric average. To describe the convergence of the geometric average and hence of $\mathrm{Tr}(\rho_{\ell}^2)$ we rewrite the geometric average as
 
\begin{equation}
 \overline{\xi}_{g} =\biggl(\prod_{\ell'=1}^{\ell}   \xi_{\ell '} \biggr)^{\frac{1}{\ell}} = \exp \biggl[ \frac{1}{\ell} \sum_{\ell'=1}^{\ell} \ln (\xi_{\ell'} ) \biggr]
\overset{\ell \gg 1}{ \rightarrow}  \langle \xi_g \rangle.
\end{equation}
 The strategy is to describe the convergence of $\ln ( \xi_{\ell})$ to its arithmetic average, and then translate this to the geometric average. However, we first need to argue that also the arithmetic average of $\ln ( \xi_{\ell})$ is independent of the layer $\ell$ in a first order approximation in the noise strength.

 First we want to show that not only $\braket{\xi_{\ell}}$ but also $ \braket{\ln(\xi_{\ell})}$  is independent of the layer $\ell$. Let $\tilde{p}$ denote a parameter that quantifies the noise strength. For weak noise we assume that $\xi_{\ell}$ is well described by a first order expansion in $\tilde{p}$ as $\xi_{\ell} \approx 1 + c_{U} \tilde{p}$, where $c_U$ can depend on the random unitary we sampled from our unitary $2$-design. Averaging over the unitary $2$-design we get

\begin{align}
    \braket{\xi_{\ell}} &\approx \braket{1 + c_{U} \tilde{p}} = 1+ \braket{ c_U } \tilde{p}  \\
    \braket{\ln(\xi_{\ell})} &\approx \braket{\ln(1 + c_U \tilde{p})} \approx \braket{ c_U} \tilde{p} . 
\end{align}
Since $\braket{\xi_{\ell}}$ is approximately independent of the layer $\ell$ as shown in Eq.~\eqref{eq:haar_one_layer}, we see that, within the same approximation, this is also true for $ \braket{\ln(\xi_{\ell})}$. Besides, the range we found for $\xi_{\ell}$ in Eq.~\eqref{eq:range_r} can be easily transferred to $\ln(\xi_{\ell})$ since the natural logarithm is a strictly monotone function. Thus, the range $R_{\mathrm{ln}}$ of the random variable $\ln(\xi_{\ell})$ can be taken as

\begin{equation}
    R_{\mathrm{ln}} = \ln(\lambda_{\mathrm{max}}) - \ln(\lambda_{\mathrm{min}}).
\end{equation} 

We now show that we can further approximate the geometric average with the arithmetic average to first order in the noise strength. In fact 

\begin{equation}
  \langle \xi_g \rangle
  =\exp \biggl[  \langle \ln( \xi ) \rangle \biggr ]
    \approx  \exp \biggl[   \langle \ln(  1 + c_{U_{\ell'}} \tilde{p} ) \rangle \biggr ]
    \approx \exp \biggl[  \langle  c_{U_{\ell'}} \tilde{p}  \rangle \biggr ]
    \approx  
      \langle   1 + c_{U_{\ell'}} \tilde{p}  \rangle
      \approx \langle \xi \rangle . \label{eq:match_geo_ar}
\end{equation}
With this observation, we can describe the convergence of the arithmetic sample average of $\ln(\xi_{\ell})$ to its true average. If the sample average of $\ln(\xi_{\ell})$ deviates by more that $\Delta$ from the true average we get the following bound on the random variable $\overline{\xi}_{g}$

\begin{equation}
    e^{\langle \ln(\xi) \rangle} e^{-\Delta} < \overline{\xi}_{g} < e^{\langle \ln(\xi) \rangle} e^{\Delta},
\end{equation}
and from Eq.~\eqref{eq:match_geo_ar}

\begin{equation}
     \langle \xi \rangle e^{-\Delta} \lesssim \overline{\xi}_{g} \lesssim \langle \xi \rangle e^{\Delta}.
\end{equation}
In terms of the purity at layer $\ell$ this means that we have an instance such that

\begin{equation}
\mathrm{Tr}(\rho_0^2) \langle \xi \rangle^{\ell} e^{-\ell \Delta}
\lesssim   \mathrm{Tr}(\rho_{\ell}^2)
\lesssim \mathrm{Tr}(\rho_0^2) \langle \xi \rangle^{\ell} e^{\ell \Delta}. 
\end{equation}
Using Eq.~\eqref{eq:pur_approx} in the main text we get

\begin{equation}
\label{eq:puritybound}
\langle  \mathrm{Tr}(\rho_{\ell}^2) \rangle e^{-\Delta \ell} 
\lesssim     \mathrm{Tr}(\rho_{\ell}^2)
\lesssim \langle  \mathrm{Tr}(\rho_{\ell}^2) \rangle  e^{ \Delta \ell}. 
\end{equation}
Applying Hoeffding's inequality Eq.~\eqref{eq:hoeffding} to the random variable $\ln(\xi_{\ell})$, we see that a deviation by $\Delta$ from the true average at layer $\ell$ happens with probability at most

 \begin{equation}
    P_{\mathrm{max}}= 2 \exp \biggl[\frac{-2 \ell \Delta^2}{R_{\ln}^2} \biggr].
 \end{equation}
Inverting for $\Delta$ we can express the purity bounds in Eq.~\eqref{eq:puritybound} explicitly in terms of $P_{\mathrm{max}}$ as

 \begin{equation}
 \label{eq:purityboundpmax}
     \langle  \mathrm{Tr}(\rho_{\ell}^2) \rangle \exp \biggl[- 
     \sqrt{\frac{1}{2} \ln \biggl(\frac{2}{P_{\mathrm{max}}} \biggr)} R_{\mathrm{ln}} \sqrt{\ell} \biggr] \lesssim     \mathrm{Tr}(\rho_{\ell}^2) \lesssim  \langle  \mathrm{Tr}(\rho_{\ell}^2) \rangle \exp \biggl[+
     \sqrt{\frac{1}{2} \ln \biggl(\frac{2}{P_{\mathrm{max}}} \biggr)} R_{\mathrm{ln}} \sqrt{\ell} \biggr].
 \end{equation}
 This means that at least $(1-P_{\mathrm{max}})  \% $ of the single circuit run purities will (approximately) lie within the derived purity bounds in a random experiment. We use Eq.~\eqref{eq:puritybound} to bound the purity in Fig.~\ref{fig:hoeffd_bounds} in the main text for the case of amplitude damping noise described in Appendix~\ref{app:subsubsecad}.

 \section{Additional numerical results for QAOA circuits}
 \label{sec:additional_qaoa}

\subsection{ Average purity decay in QAOA circuits with more qubits} 
\label{appsubsec:more_purs_qaoa}

In this section, we give additional numerical results for the decay of the average purity in QAOA circuits under amplitude damping noise. In particular, we consider circuits with 6, 8, 10 and 12 qubits up to 40 layers\footnote{We used a smaller circuit depth than in Fig.~\ref{fig:pur_qaoa} for computational reasons.}. The circuits correspond to 3-regular graphs and Erdos-Renyi graphs. To create an Erdos-Renyi graph one needs to fix the number of vertices and the probability to create an edge between 2 vertices. The Erdos-Renyi graph is then created randomly. We fix the probability of creating an edge to $0.5$. We use the same 6-vertex 3-regular graph as for Fig.~\ref{fig:pur_qaoa} in the main text, which is shown in Fig.~\ref{fig:used_graphs}. The other 3-regular and Erdos-Renyi graphs that we use are shown in Fig.~\ref{fig:used_graphs_more_qubits}.

\begin{figure}[htb]
    \centering
    \includegraphics[width=\textwidth]{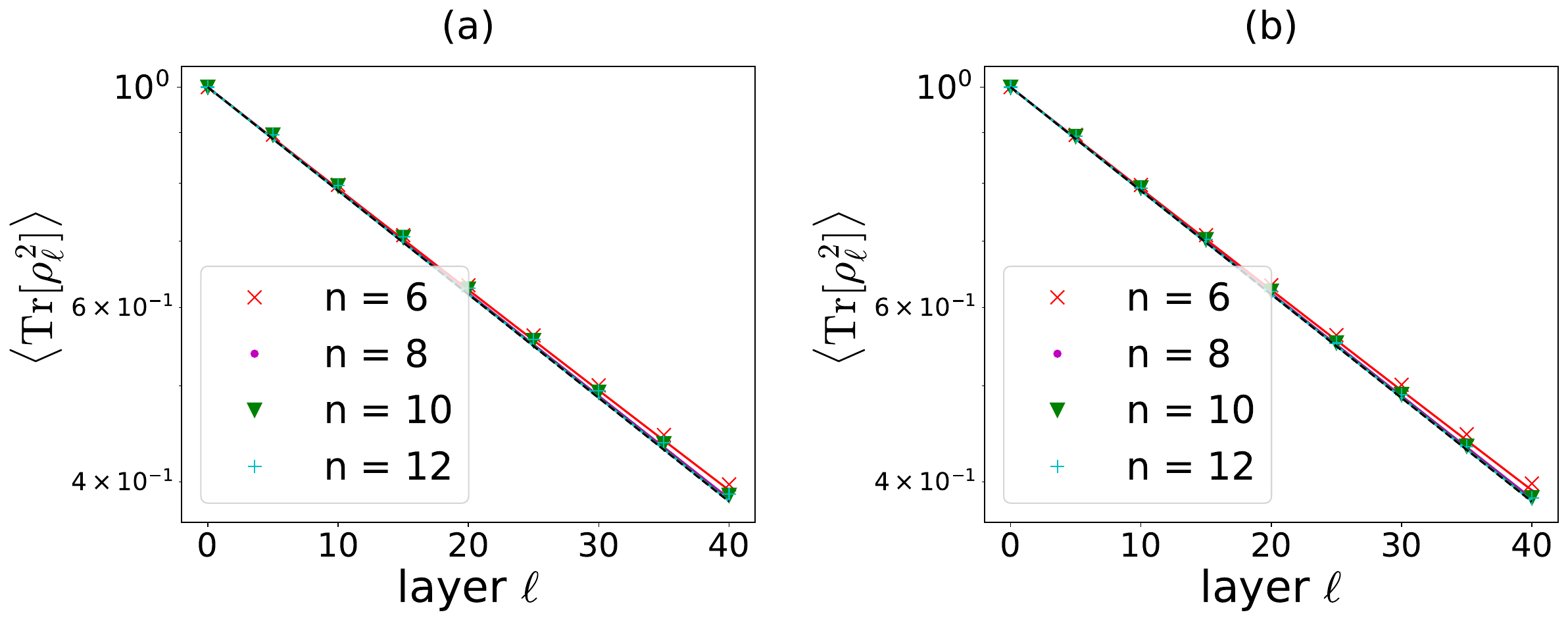}
    \caption{  Average purity decay for  different QAOA circuits with qubit number $n$ under amplitude damping noise. The circuits used in (a) correspond to 3-regular graphs. The circuits used in (b) correspond to Erdos-Renyi graphs. For the Erdos-Renyi graphs, we fix the probability of edge creation to 0.5. We get the average purity by evaluation of 128 samples with random circuit parameters that are drawn from the uniform distribution $\alpha_{\ell}, \gamma_{\ell} \in [0, 2 \pi]$. We choose the local amplitude damping noise strength $\gamma_{\downarrow}$ such that the effective depolarizing strength $p_{\mathrm{eff}}$ of the channel $\mathcal{N}_{\mathrm{depol}}(\rho) = (1-p_{\mathrm{eff}}) \rho +p_{\mathrm{eff}} \frac{I}{2^n} $ is the same for every graph. Here $n$ represents the number of qubits. In particular, we use $p_{eff} \approx 0.012$. For 6 qubits this means $\gamma_{\downarrow}=0.004$, which is the noise strength corresponding to the red curve in Fig.~\ref{fig:pur_qaoa} of our paper. The different colors stand for the different graphs. The symbols indicate the purity decay in the QAOA circuit and the solid lines give the analytical exact average purity decay. Since $p_{\mathrm{eff}}$ is the same for all graphs, the analytical approximate average purity decay is the same for all graphs and is represented by the dashed black line.
     }
    \label{fig:3_reg_Renyi}
\end{figure}

Our analytical prediction of the average purity assumes weak noise. Thus, to make a fair comparison between different numbers of qubits, we keep the effective depolarizing strength $p_{\mathrm{eff}}$ constant. This means that we need to decrease the local amplitude damping noise strength $\gamma_{\downarrow}$ as we increase the number of qubits. In particular we fix $p_{\mathrm{eff}} \approx 0.012$. For 6 qubits this means $\gamma_{\downarrow}=0.004$. This is the noise strength corresponding to the red curve of Fig.~\ref{fig:pur_qaoa} in the main text. We show the resulting purities for the 3-regular graphs in Fig.~\ref{fig:3_reg_Renyi}(a) and the purities for the Erdos-Renyi graphs in Fig.~\ref{fig:3_reg_Renyi}(b).

The black dashed line represents the analytical approximate average purity decrease. This decrease is the same for all graphs as we keep $p_{\mathrm{eff}}$ constant. The purities corresponding to the QAOA circuits are indicated by symbols, and the solid lines give the analytical exact average purity decrease. We see that for both types of graphs, the 3-regular graphs and the Erdos-Renyi graphs, the purities from the QAOA circuits are close to the analytical predictions. Besides, we see that with increasing qubit number the approximate average purity gets closer to the exact average purity. This can be explained by observing that to get the analytical approximate average purity in Subsec.~\ref{subsec:puritydecay}, we assumed that the purity is much larger than $1/2^n$, and this assumption is better satisfied for large qubit numbers.

 \subsection{Twirling with QAOA circuits} 
\label{appsubsec:twirlingqaoa}

In this section, we want to get a qualitative understanding of how well the behavior of the different QAOA circuits under amplitude damping noise is captured by the toy model. To proceed we study if the QAOA circuits with initial state $\ket{+}^{\otimes n}$ twirl amplitude damping noise according to the defining equation Eq.~\eqref{eq:twirling_eq} of unitary $2$-designs. Twirling noise with a unitary $2$-design results in a global depolarizing channel. To see if this property approximately holds for a given QAOA circuit, we consider the fidelity between the states 

\begin{subequations}
\label{eq:qaoahaarstates}
    \begin{equation}
     \mathcal{A}_{\mathrm{QAOA}}(\ket{+}\bra{+}^{\otimes n}) = 
     \int_{\mathcal{D}} d \bm{\alpha} d \bm{\gamma}  \mathcal{U}^{\dagger}(\bm{\alpha}, \bm{\gamma}) \circ \mathcal{A}^{\otimes n} \circ \mathcal{U}(\bm{\alpha}, \bm{\gamma}) (\ket{+}\bra{+}^{\otimes n})
     \end{equation}

\begin{equation}
     \mathcal{A}_{\mathrm{Haar}}(\ket{+}\bra{+}^{\otimes n}) = \\
     (1-p_{\mathrm{eff}}) \ket{+}\bra{+}^{\otimes n} + p_{\mathrm{eff}} \frac{I}{2^n}
\end{equation}
\end{subequations}
We are interested in evaluating the quantum state fidelity between $\mathcal{A}_{\mathrm{QAOA}}(\ket{+}\bra{+}^{\otimes n})$ and $\mathcal{A}_{\mathrm{Haar}}(\ket{+}\bra{+}^{\otimes n})$ defined for any $\rho, \sigma \in \mathrm{D}_n$ as $F(\rho, \sigma)=\mathrm{Tr}(\sqrt{\sqrt{\rho} \sigma \sqrt{\rho}}\bigr)$. If the QAOA circuit approximates well the behavior of a unitary $2$-design with respect to amplitude damping noise, we expect this fidelity to be close to $1$. Note that this does not mean that the QAOA circuit forms a unitary $2$-design, since we are analyzing only amplitude damping and the initial state $\ket{+}\bra{+}^{\otimes n}$. A proper test of the unitary $2$-design property can be numerically quite expensive \cite{cleve2016, kim2019, Nakaji2021} and it is beyond the scope of our analysis. In fact, in Subsec.~\ref{subsec:pur_QAOA} we find for the considered circuits that the larger the fidelity, the better the purity predicted by the toy model in Subsec.~\ref{subsec:nibp2design} matches the purity of the QAOA circuit. We compute the fidelity between the states in Eq~\eqref{eq:qaoahaarstates} as a function of the number of QAOA layers for randomly chosen $6$-vertex $5$-regular, $4$-regular, $3$-regular and $2$-regular graphs. The chosen graphs can be found in Fig.~\ref{fig:used_graphs} in App.~\ref{app:used graphs}. Besides, we consider a QAOA circuit that is known to become universal at a certain depth, and therefore can be seen as a perfect unitary $2$-design. The universal circuit was originally proposed in Ref.\:\cite{lloyd} and further analyzed in Ref.\:\cite{universality}. We give details about its implementation in App.~\ref{sec:app_uni}. Note that for the universal QAOA circuit we need an odd number of qubits and we choose 5 qubits, while the MaxCut circuits contain 6 qubits. Therefore, we choose the noise strength in the universal QAOA circuit such that we get the same effective depolarizing strength $p_{\mathrm{eff}}$ as for the MaxCut circuits with 6 qubits.

\begin{figure}
    \centering
    \includegraphics[width=0.6\textwidth]{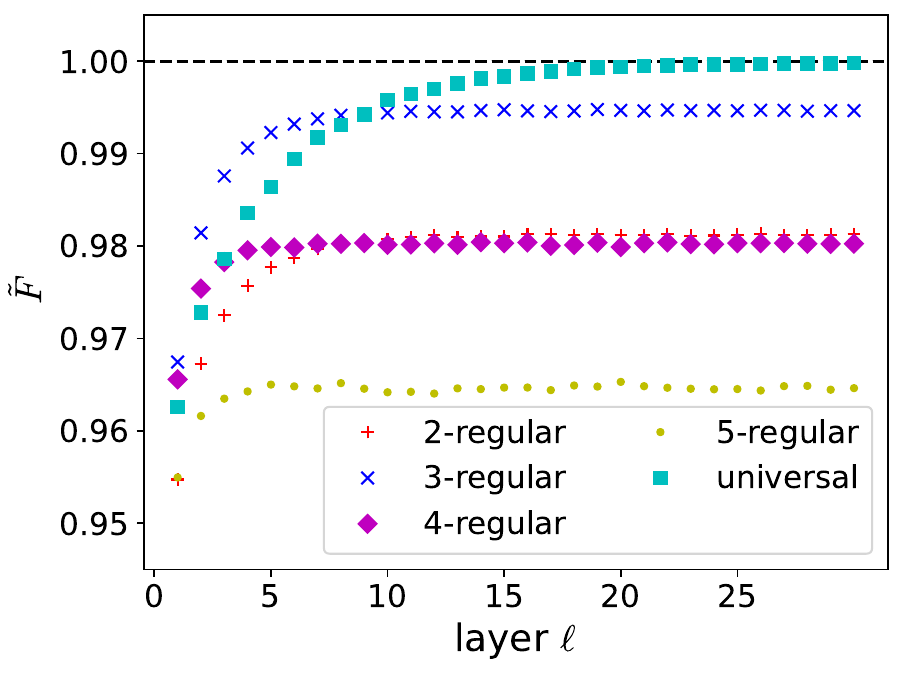}
    \caption{Twirling the amplitude damping channel with QAOA circuits for MaxCut problems on a 6-vertex $d$-regular graph. The parameter of amplitude damping $\gamma_{\downarrow}$ is set to $\gamma_{\downarrow}= 0.06$. The universal QAOA circuit, described in Appendix~\ref{sec:app_uni}, is considered for $5$ qubits and with $\gamma_{\downarrow} = 0.072$. For each circuit, we take $768$ sample, with parameters that are chosen randomly from the uniform distribution $\alpha_{\ell},\gamma_{\ell} \in [0,2 \pi )$ to obtain $\mathcal{A}_{\mathrm{QAOA}}(\ket{+}\bra{+}^{\otimes n}) $. The additional parameters needed to define the universal QAOA circuit are taken as $\omega_{A}\approx 0.45$, $\omega_{B}\approx 0.54$, $\gamma_{AB}\approx 0.22$, $\gamma_{BA}\approx 0.52$, in order to satisfy the conditions in Eq.~~\eqref{eq:cond} in Appendix~\ref{sec:app_uni}. The fidelity $\tilde{F} = F (  \mathcal{A}_{\mathrm{QAOA}}(\ket{+}\bra{+}^{\otimes n}) ,\mathcal{A}_{\mathrm{Haar}}(\ket{+}\bra{+}^{\otimes n})   )$ increases with increasing circuit depth as expected for any graph, but it only approaches $1$ for the universal QAOA circuit.}
    \label{fig:twirling_all_AD_006}
\end{figure}

The resulting fidelities are shown in Fig.~\ref{fig:twirling_all_AD_006}. We see that for all circuits the fidelity increases with an increasing number of layers. This is because, as the circuit depth grows, the circuits can reach more states. The only circuit that reaches $F(\mathcal{A}_{\mathrm{QAOA}}(\ket{+}\bra{+}^{\otimes n})),\mathcal{A}_{\mathrm{Haar}}(\ket{+}\bra{+}^{\otimes n}))=1$ is the universal circuit. This is due to the fact that QAOA circuits for MaxCut do not form a perfect unitary $2$-design. We see that how well a QAOA circuit for a MaxCut problem twirls the noise depends on the underlying graph. From our choice of graphs the $6$-vertex $3$-regular graph is the one that twirls the noise best, while the $6$-vertex $5$-regular graph twirls the noise worst.

\subsection{How well is the noise modelled by $\mathcal{A}_{\mathrm{Haar}}$?} \label{sec:noise_channel}

\begin{figure}
    \centering
    \includegraphics[width=1\textwidth]{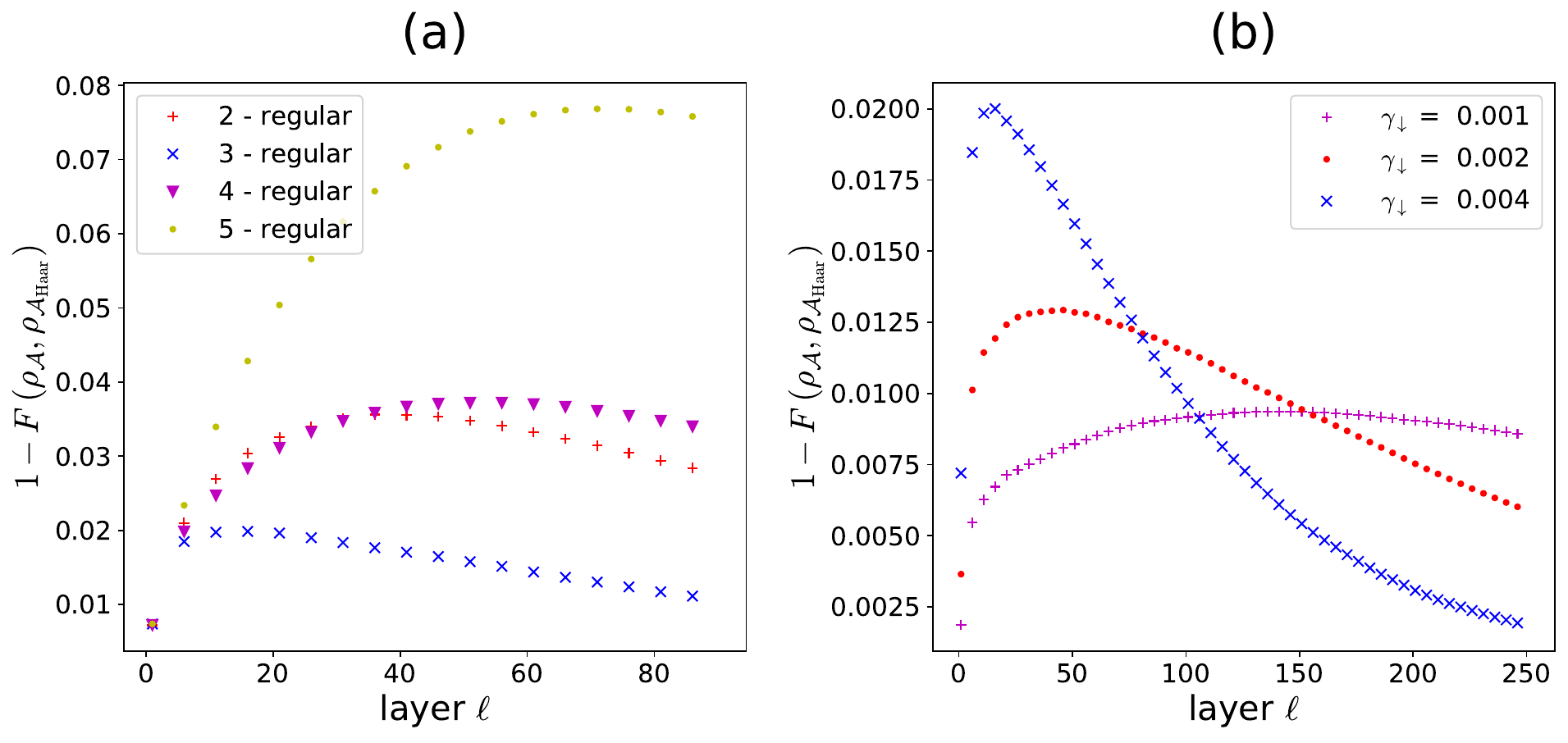}
    \caption{Average infidelity between the states $\rho_{\mathcal{A}}(\bm{\alpha}, \bm{\gamma})$, $\rho_{\mathcal{A}_{\mathrm{Haar}}}(\bm{\alpha}, \bm{\gamma})$ as a function of the number of layers. The average infidelity is computed at each point by taking $64$ random QAOA parameters samples from the uniform distribution $\alpha_{\ell},\gamma_{\ell} \in [0,2 \pi )$. (a) Average infidelity for the $d$-regular graphs in Fig.~\ref{fig:used_graphs} with amplitude damping parameter $\gamma_{\downarrow}=0.004$. (b) Average infidelity for the $3$-regular graph in Fig.~\ref{fig:used_graphs} for different noise strengths.}
    \label{fig:infids}
\end{figure}

In Subsec.~\ref{subsec:pur_QAOA} we found that for amplitude damping noise, the average purity in our QAOA circuits is well described by the approximate average purity of the toy model of Subsec.~\ref{subsec:nibp2design}. The approximate average purity of the toy model at layer $\ell$ coincides with the purity obtained by applying $\ell$ times the global depolarizing channel $\mathcal{A}_{\mathrm{Haar}}$ to the initial state. Thus, $\mathcal{A}_{\mathrm{Haar}}$ captures the behavior of the average purity in our QAOA circuits. We now want to go one step further and study whether the noise channel in the circuit can be described by $\mathcal{A}_{\mathrm{Haar}}$ for larger number of layers. While being a conceptually interesting question this might also be of practical importance for the implementation of error mitigation techniques. In fact, knowing that the noise can be replaced by an effective global depolarizing channel makes it easily invertible via probabilistic error cancellation \cite{temme2017}.

To study this question, we randomly sample parameters $\bm{\alpha}, \bm{\gamma}$ in QAOA circuits and compute the final state under amplitude damping noise. We denote this state by $\rho_{\mathcal{A}}(\bm{\alpha}, \bm{\gamma})$. For the same parameters, we replace each amplitude damping channel with the corresponding $\mathcal{A}_{\mathrm{Haar}}$  and obtain the state $\rho_{\mathcal{A}_{\mathrm{Haar}}}(\bm{\alpha}, \bm{\gamma})$. If $\mathcal{A}_{\mathrm{Haar}}$ is a good description of the noise process the two resulting states should be close to each other. We study this by computing the infidelity $1- F(\rho_{\mathcal{A}}(\bm{\alpha}, \bm{\gamma}),\rho_{\mathcal{A}_{\mathrm{Haar}}}(\bm{\alpha}, \bm{\gamma}))$. In Fig.~\ref{fig:infids}(a) we plot the average infidelities for the different $6$-vertex $d$-regular graphs in Fig.~\ref{fig:used_graphs}. In all cases the infidelity first increases with increasing circuit depth before it decreases again. Comparing the results to Fig.~\ref{fig:twirling_all_AD_006}, we see that the circuits that twirl better amplitude damping noise into $\mathcal{A}_{\mathrm{Haar}}$, result in lower maximal infidelities. This suggests that in these circuits the error made by modelling the noise as $\mathcal{A}_{\mathrm{Haar}}$ is smaller. Besides, there is a difference in the position of the maximal infidelity for the considered circuits. Circuits that do not twirl amplitude damping noise well into $\mathcal{A}_{\mathrm{Haar}}$   tend to have their maximal infidelity at a higher depth. In Fig.~\ref{fig:infids}(b), we consider the $3$-regular graph for different noise strengths. We see that decreasing the noise strength decreases the maximal infidelity. The reason for this might be that if the local noise strength is smaller, we apply less noise before the circuit reaches its maximal ability to twirl amplitude damping noise into $\mathcal{A}_{\mathrm{Haar}}$. 

\section{Universal QAOA circuit}
\label{sec:app_uni}
\begin{figure}
    \centering
    \includegraphics[width=0.6\textwidth]{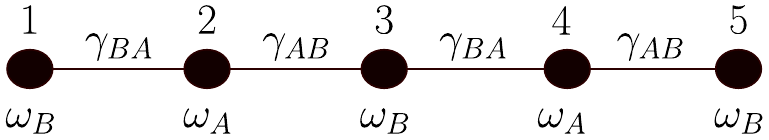}
    \caption{Representation of the problem Hamiltonian in the universal QAOA circuit in the $5$-qubit case. The problem Hamiltonian can be associated with a $1$-dimensional chain.}
    \label{fig:lloyd_circuit}
\end{figure}

Here we introduce the universal QAOA circuit, proposed in Ref.\:\cite{lloyd} and further analyzed in Ref.\:\cite{universality}. It is defined by

\begin{align}
    H_M &= \sum_{j} X_{j} \label{lloyd_hm}, \\
    H_P &= \sum_{j} \omega_{A} Z_{2j} + \omega_{B} Z_{2j+1}  + \gamma_{AB} Z_{2j} Z_{2j+1} +\gamma_{BA} Z_{2j+1} Z_{2j+2}, \label{lloyd_hp}
\end{align}
where $H_{\mathrm{P}}$ is acting on a one-dimensional line as shown in Fig.~\eqref{fig:lloyd_circuit} for a $5$-qubit example. In Ref.\:\cite{universality} it is shown that this circuit is only universal if we have an odd number of qubits and besides the following conditions are satisfied

\begin{align}
    \omega_{A}^2 &\neq \omega_{B}^2 \nonumber \\
    \gamma_{AB}^2 &\neq \gamma_{BA}^2 \nonumber \\
    \gamma_{AB}^2 - 4 \gamma_{BA}^2 &\neq 0 \nonumber \\
    \gamma_{AB} &\neq 0  \nonumber \\
    \gamma_{BA} &\neq 0. \label{eq:cond}
\end{align}

\section{Used graphs} 
\label{app:used graphs}

\begin{figure}
    \centering
    \includegraphics[width=\textwidth]{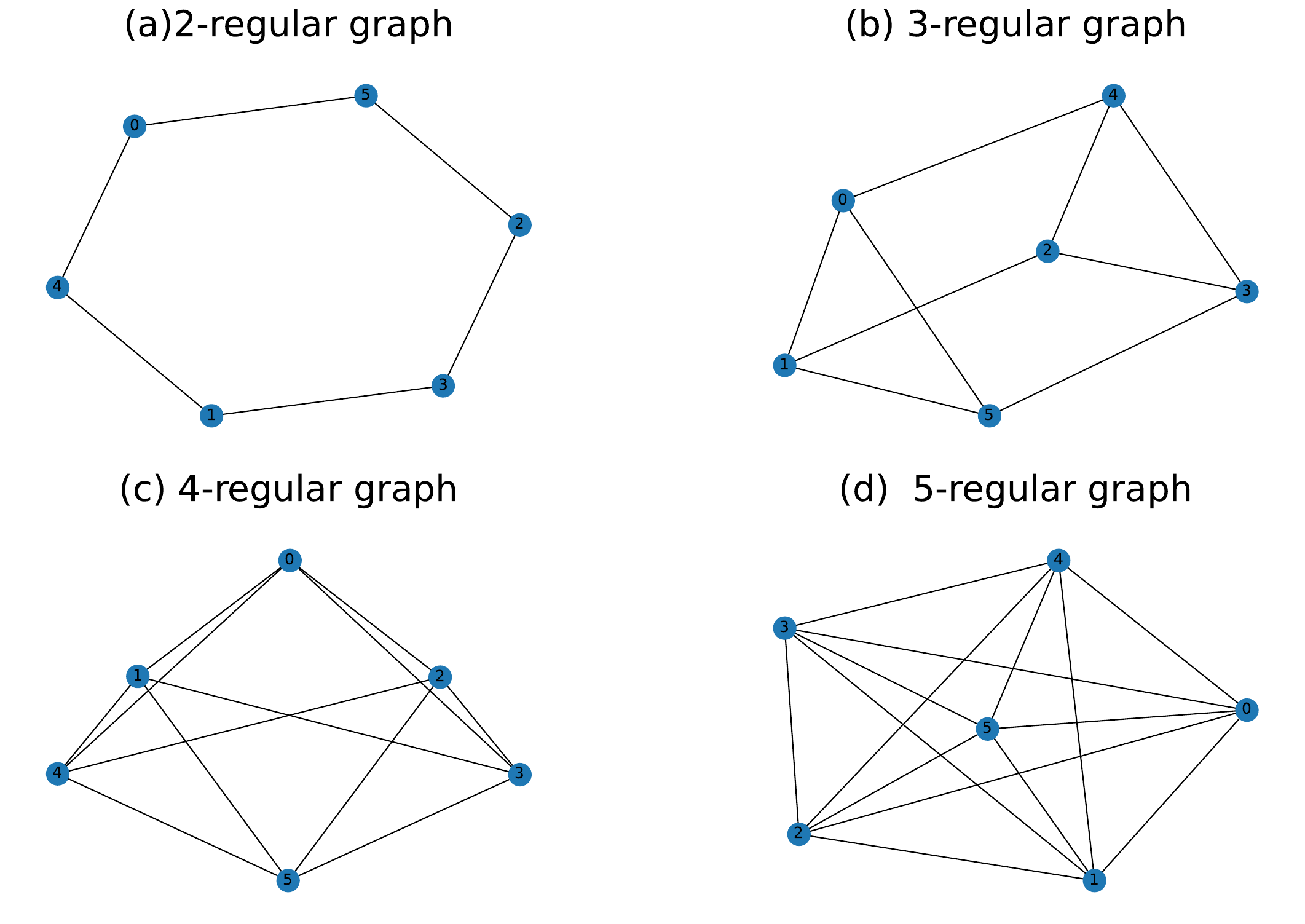}
    \caption{The $6$-vertex $d$-regular graphs used in our numerical analysis discussed in Sec.~\ref{sec:numerical} in the main text.}
    \label{fig:used_graphs}
\end{figure}

\begin{figure}
    \centering
    \includegraphics[width=\textwidth]{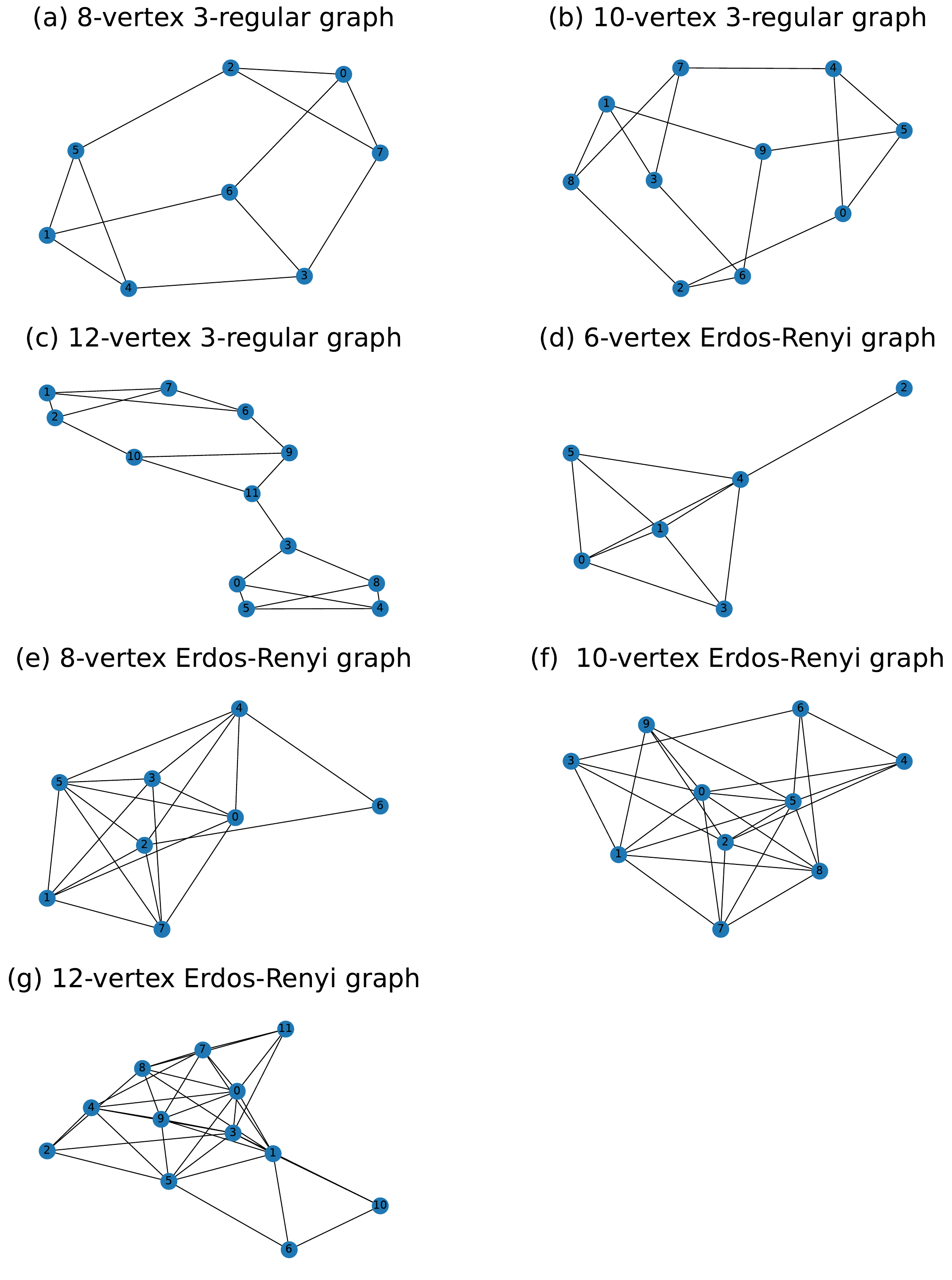}
    \caption{Different 3-regular and Erdos-Renyi graphs used in our numerical simulations. We analyze the purity decay of the corresponding QAOA circuits in App.~\ref{appsubsec:more_purs_qaoa}.}
    \label{fig:used_graphs_more_qubits}
\end{figure}

In the main section, we run QAOA circuits for MaxCut problems on randomly chosen $6$-vertex $d$-regular graphs. In Fig.~\ref{fig:used_graphs} we show the graphs we used. In App.~\ref{appsubsec:more_purs_qaoa}, we provide additional numerical results for circuits with a bigger qubit number. The corresponding graphs can be seen in Fig.~\ref{fig:used_graphs_more_qubits}.

\end{document}